\documentclass[preprint,3p,times,sort&compress]{elsarticle}

\usepackage{multirow,setspace,times,amssymb,amsmath,graphicx,color,rotating,subfigure,url}
\usepackage{lineno}
\usepackage{natbib}
\usepackage{booktabs}%
\usepackage{longtable}%
\graphicspath{{Figures/}}
\usepackage{subfigure}
\usepackage[table]{xcolor}
\usepackage{tabularx}
\usepackage{graphicx} 
\usepackage{epstopdf}
\usepackage{mathrsfs}
\usepackage{makecell}
\usepackage[bookmarks=true,colorlinks,linkcolor=blue,anchorcolor=blue,citecolor=blue,unicode]{hyperref}
\usepackage{bookmark}

\journal{American Journal of Agricultural Economics}
\graphicspath{{Figures/}}
\begin{document}

\begin{frontmatter}
\newpage
\title{Spatiotemporal characteristics of agricultural food import shocks}

\author[SB,RCE]{Yin-Ting Zhang}
\author[IPAG,VNU]{Duc Khuong Nguyen}
\author[SB,RCE,Math]{Wei-Xing Zhou\corref{cor1}}
\ead{wxzhou@ecust.edu.cn} 
\cortext[cor1]{Corresponding author.}
\address[SB]{School of Business, East China University of Science and Technology, Shanghai 200237, China}
\address[RCE]{Research Center for Econophysics, East China University of Science and Technology, Shanghai 200237, China}
\address[IPAG]{IPAG Business School, Paris, France}
\address[VNU]{International School, Vietnam  National University, Hanoi, Vietnam}
\address[Math]{School of Mathematics, East China University of Science and Technology, Shanghai 200237, China}

\begin{abstract}
Ensuring food supply stability is key to food security for economies, and food imports become increasingly important to safeguard food supplies in economies with inadequate food production. Food import shocks have significant impacts on targeted economies. Using import trade data of four staple crops (maize, rice, soybean, and wheat) from 1986 to 2018, this paper identifies food import trade shocks that occurred to economies during the period of 1995--2018. We compare the temporal evolution and spatial distribution of import shocks occurring to different crops and analyze the shock intensity and shock recovery in various continents based on locally weighted polynomial regression and Cook's distance. The results reveal higher frequencies during the 2007/2008 food crisis and relatively higher shock frequencies in North America, Africa, and Asia. Meanwhile, there are regional differences in shock recovery, with the majority of shocks in Asia recovering in the short term. We also find that high import diversity and a low import dependency ratio buffer economies against import shocks, resulting in a low shock rate and a high recovery rate. These results contribute to our understanding of the external supply risks of food, placing emphasis on accessibility issues in food security.
\end{abstract}
	
\begin{keyword} 
Econophysics; Agricultural food import; Shock identification; Recovery; Food security
\\
  JEL: C1, P4, Z13
\end{keyword}

\end{frontmatter}


\section{Introduction}
Food security has traditionally been placed a high importance by economies all over the world and has once again come to the top of the agenda \cite{Zhan-Zhang-Li-Zhang-Qi-2020-AnnOperRes}. In 2020, the COVID-19 pandemic not only has a negative effect on the global economy \cite{,Mutaju-2022-CFRI,Sajid-Ahmad-2022-CFRI}, but also causes a rise in the number of hungry people in the world by destroying food supply chains \cite{Sharma-Singh-Kumar-Mani-Venkatesh-2021-AnnOperRes}. The incidence of undernourishment rose from 8.4\% to 9.9\% in just one year after holding stable for five years, making the aim of ending hunger by 2030 even more challenging\footnote{The State of Food Security and Nutrition in the World (2021), available at {\url{https://www.fao.org}}. Accessed 21 June 2022.}. The 2022 Global Report on Food Crises recently published alerts that the 193 million people were facing acute food insecurity, which continues to escalate\footnote{Hunger Hotspots-FAO-WFP early warnings on acute food insecurity (2022), available at {\url{https://www.fao.org/3/cc0364en/cc0364en.pdf}}. Accessed 21 June 2022.}. The COVID-19 is served as a stark reminder that the advancements in food security we have attained are merely transitory successes. A crucial indicator of food security is food availability. Global food availability has changed dramatically in the past few decades \cite{Porkka-Kummu-Siebert-Varis-2013-PLoSOne}. Conflicts \cite{Jagermeyr-Robock-Elliott-Muller-Xia-Khabarov-Folberth-Schmid-Liu-Zabel-Rabin-Puma-Heslin-Franke-Foster-Asseng-Bardeen-Toon-Rosenzweig-2020-ProcNatlAcadSciUSA}, climatic variability and climate extremes \cite{Tilman-Balzer-Hill-Befort-2011-ProcNatlAcadSciUSA}, and economic slowdown and recession have made it increasingly difficult for agricultural production to sustain the growing population. Besides, there is also a large amount of food waste in some areas \cite{Mogale-Kumar-Tiwari-2020-AnnOperRes,Dora-Wesana-Gellynck-Seth-Dey-DeSteur-2020-AnnOperRes}, which leaves a shortage of food supplies to fulfill local demands. Food is harvested both by local farms and isolated markets through a network of interactions (for example, trade) 
\cite{Tu-Suweis-D'Odorico-2019-NatSustain}.

The development of food packaging technology and the reduction of transportation costs have enhanced efficient global food distribution through food trade. Food imports play an important role in making up for the domestic market at times of adverse domestic supply shocks \cite{Chen-Villoria-2019-EnvironResLett}. Some economies cannot meet people's food demand through local production and need to address food self-sufficiency through food imports \cite{Garre-Fernandez-Brereton-Elliott-Mojtahed-2019-FoodResInt}. At present, at least 34 economies are unable to produce adequate food due to a lack of water and land resources, resulting in a part of the population that relies on imported food to avoid starvation. Economies such as Afghanistan, the Central African Republic, the Democratic Republic of the Congo, Djibouti, and Ethiopia have long been facing food crises due to insufficient domestic food production\footnote{Countries Most Dependent On Others For Food (2017), available at {\url{https://www.worldatlas.com/articles/the-countries-importing-the-most-food-in-the-world.html}}. Accessed 29 November 2022.}. These economies can only rely on food imports and international food aid to meet basic food consumption. Once food imports are blocked, they will face a huge food supply crisis. The disruption of food supply chains resulting from the conflict between Russia and Ukraine in 2022 and restrictions on Russian exports will have a significant impact on food security. Russia and Ukraine play a major part in global food production and supply. Approximately 50 economies currently rely on imports from Russia and Ukraine to secure 30\% or more of their wheat supply, and most of them are least developed economies or low-income food-deficit economies in North Africa, Asia and the Near East\footnote{New Scenarios on Global Food Security based on Russia-Ukraine Conflict (FAO), available at {\url{https://www.fao.org/director-general/news/news-article/zh/c/1476506/}}. Accessed 21 June 2022.}. The situation with regard to food security is particularly dire for these economies. Thus, food import trade soared in importance in guaranteeing food supply, both as a source of long-term food supply \cite{Porkka-Kummu-Siebert-Varis-2013-PLoSOne,Porkka-Guillaume-Siebert-Schaphoff-Kummu-2017-EarthFuture}, and also as an emergency supplement to short-term food shortages caused by climate changes \cite{Baldos-Hertel-2015-FoodSecur}.

International food import and export trade boosts the availability of food by connecting surplus economies and deficit economies \cite{Baldos-Hertel-2015-FoodSecur}, and strengthens the stability of the food system. On the other hand, the risks of food shortage and price rise are spread through the international food trade network, which exacerbates food insecurity\footnote{Evaluation of the USDA Fruit and Vegetable Pilot Program: Report to Congress (2003), available at {\url{https://www.ers.usda.gov/publications/pub-details/?pubid=43277}}. Accessed 29 November 2022.}. When trade links between economies are tighter, the contagion of shocks becomes wider \cite{Foti-Pauls-Rockmore-2013-JEconDynControl}. High import-dependent economies are particularly vulnerable to exporting economies \cite{Kummu-Kinnunen-Lehikoinen-Porkka-Queiroz-Roos-Troell-Well-2020-GlobFoodSecur-AgricPolicy}. Insufficient production in exporting economies respond by reducing or restricting exports, and rising global food prices have further exacerbated the food crisis \cite{Giordani-Rocha-Ruta-2016-JIntEcon}, affecting food imports in importing economies and causing changes in food supply in economies. We consider an outlier in import trade volumes as a shock to import trade. It is significantly lower than normal levels. The pressure to guarantee food security is increased when economies drastically reduce their imports of food.  Therefore, identifying food import trade disruptions or shocks and quantifying food external supply security is helpful to discuss the global food security from the dimension of food availability.

As a result of the sharp increases in food prices in 2007–08 and 2010–11, as well as the Arab Spring in 2010–11, more and more academics are focusing on the role that trade plays in the global food system. People have paid more attention to the transmission mechanism of price risks in the international food market \cite{Ceballos-Hernandez-Minot-Robles-2017-WorldDev}, the cascading effects of local shocks through the food supply chain \cite{Moser-Hart-2015-ClimChange}, and the blocking of key nodes in the food trade network to increase the vulnerability of the food trade system \cite{Foti-Pauls-Rockmore-2013-JEconDynControl,Wellesley-Preston-Lehne-Bailey-2017-ResTranspBusManag}. Shocks to the food import trade can limit external access to food.

In this study, we characterize the temporal and spatial properties of international food import trade shocks. Four crops (maize, rice, soybean and wheat) that provide the majority of human caloric needs are selected to construct an international food trade network using data of food trade volume from 1986 to 2018. We detect shocks through outliers in a time series' autocorrelation structure. To identify the outliers, we fit the food import time series by using the locally weighted polynomial regression (LOWESS). We determine the shock magnitude (intensity) as the difference between the abnormal value and the average value, and define the shock recovery as the 95\% level from the food trade volume to the average value before the shock. More specifically, this paper considers the import trade of different crops and compares the single food import trade with the aggregate food import trade, summing up four types of food. Through the above work, this paper aims to address the following problems: 1) Are there temporal and spatial differences in food import trade shocks? 2) Are there any differences between the four types of food import trade shocks? 3) Do shocks in a single crop import trade network have unique properties compared with shocks in aggregate crop import trade networks? To this end, we also attempt to analyze the factors that cause the temporal and spatial differences in shock frequency (the shock frequency of a given food is defined as the ratio of the number of economies experiencing shocks to the number of economies participating in food import trade in that year) and recovery time. Of course, this is only a very basic discussion. More in-depth research needs to introduce more variables and methods to explore the internal causal relationship in the future.


This paper starts with the availability analysis of food in the economy and identifies import shocks to the external food supply by testing the abnormal values of food import trade. At the same time, the shock magnitudes and recovery times are calculated to measure the intensity of food import shocks and the sustainability of food supply. Evaluating food security is a multi-dimensional, comprehensive undertaking \cite{Hubbard-Hubbard-2013-FoodPolicy}. The work of this paper is biased, but it can nonetheless help us better comprehend global food security. The rest of this article is arranged as follows. Section~\ref{S2:DataMethod} describes the data source and the methods adopted. Section~\ref{S3:Results} presents the identification results of food import trade shocks, including the spatial evolution of shocks to the four crop imports, the regional differences of shock recovery and recovery time, and the comparison between the single and aggregate food import trade shocks. Section~\ref{S4:Discussion} summarizes the results and explores the reasons for the temporal and spatial differences of food import trade shocks.

\section{Data and methods}
\label{S2:DataMethod}

\subsection{Data description}

Maize, rice, soybean and wheat are the four most important crops in the world, which meet the energy and nutrition needs of the vast majority of the world population \cite{Burkholz-Schweitzer-2019-EnvironResLett}. To characterize the evolution of international food trade structure and identify trade shocks among economies, four temporal international food trade networks are constructed from 1986 to 2018, using the data retrieved from the Food and Agriculture Organization of the United Nations (FAO). Due to the differences in import and export data reported by each economy, we use the import trade volume as the statistical standard to improve the accuracy and consistency of the data. In view of the lack of data in some small economies, the export data reported by exporting economies are used to supplement the data to ensure the integrity of the data. Each economy $i$ is a node in the international food trade network, and the link weight $w_{ij}$ means trade flows from the economy $i$ to the economy $j$.

\begin{figure}[h!]
    \centering
     \includegraphics[width=0.53\linewidth]{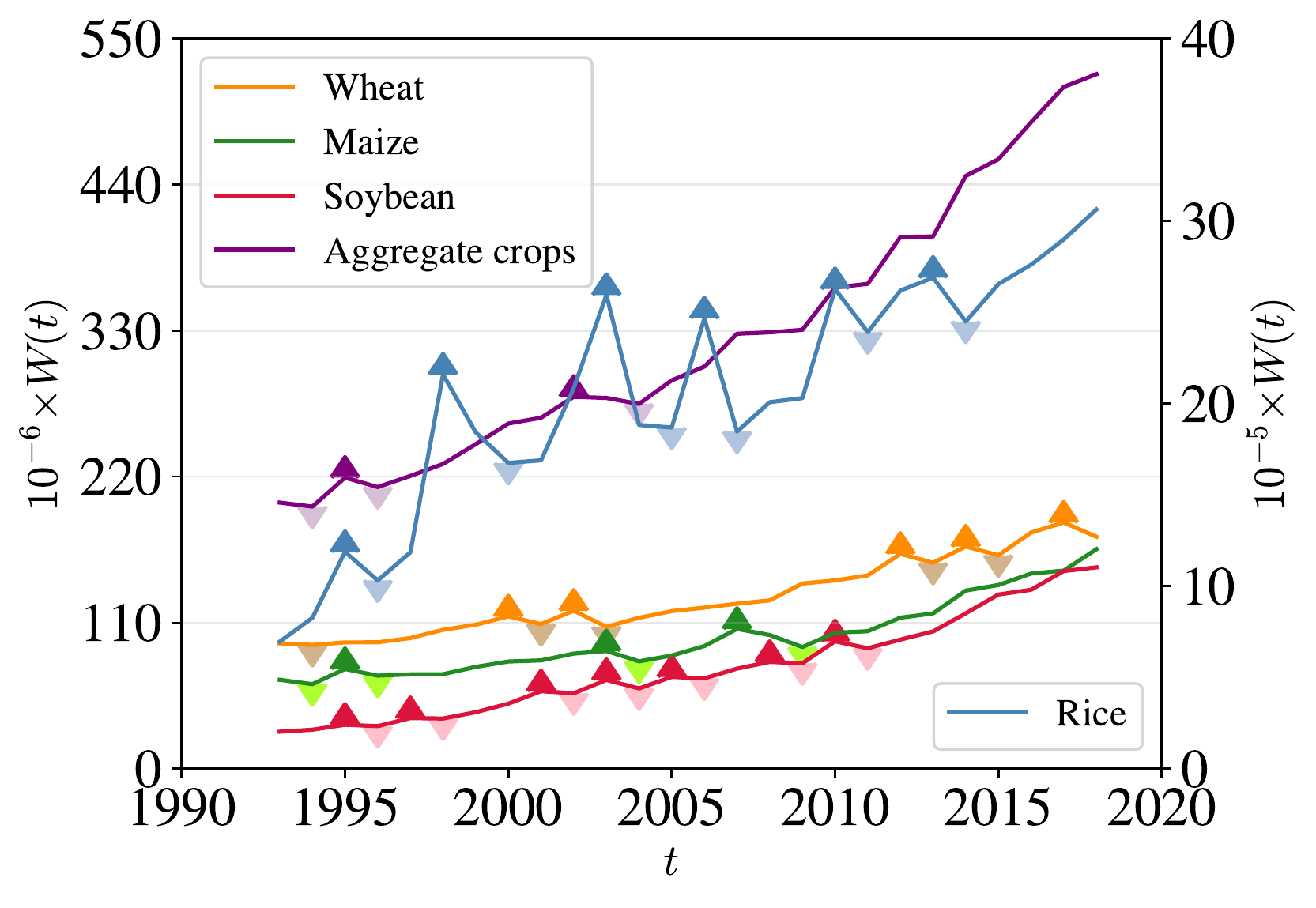}

    \caption{Total link weight $W$ in units of Tonnes. The left coordinate corresponds to wheat, maize, soybean and aggregate of four crops and the right coordinate corresponds to rice.}
    \label{Fig:iCTN:sumW:t:1986-2018}
\end{figure}

Figure~\ref{Fig:iCTN:sumW:t:1986-2018} shows the evolution of the total weight of each food trade network and the total weight of the aggregate food trade network during 1986-2018. The dark and light arrows indicate the peaks and troughs of the waves respectively. As can be seen from Figure~\ref{Fig:iCTN:sumW:t:1986-2018}, the total trade volume of all kinds of crops showed an overall upward trend during 1986-2018. The evolution of different international food trade networks shows different characteristics. The total trade volume of wheat is always the largest while the total trade volume of rice is always significantly less than the other three crops. And the total trade volume of soybean has the most obvious change trend. In general, the size of the international food trade network has expanded, and the total amount of food trade between economies has increased. We note that the weight decreased significantly in some years, which is however not sufficient to claim that food trade experienced shocks. Therefore, we used a quantitative approach to identify shocks based on outlier detection.

\subsection{Food import shock detection}
\label{S2:ImportShock:DetectionMethod}

There are many methods that can be adopted to detect the shocks of international food import trade. The quantitative approach through identifying time series outliers has been used to detecting shocks to food production and trade. Based on the method of Cottrell et al. \cite{Cottrell-Nash-Halpern-Remenyi-Corney-Fleming-Fulton-Hornborg-Johne-Watson-Blanchard-2019-NatSustain}, this paper defines shocks as breaks in the autocorrelation structure of a time series. The shocks can be characterized the frequency, the intensity, and the time to recovery. When identifying food trade shocks, the annual import supply $s_{i,food}^{\mathrm{in}}(t)$ for each food ($food =$ maize, rice, soybeans, and wheat) of an economy $i$ has 33-year observations. For simplicity, we may omit the subscript $food$ and the superscript ${\mathrm{in}}$ and use $
Imp(t)$ to represent an economy's food import trade in year $t$, where $t = \{1986, 1987, ..., 2018\}$ with the length being $N_{\mathrm{Origin}}=2018-1986+1=33$. 

For each economy $i$, we use a method called locally weighted polynomial regression (LOWESS) to remove the temporal trend of $Imp(t)$:
\begin{equation}
   Imp(t) = g(t)+\varepsilon_{t},
\end{equation}
where $g$ is a polynomial smoothing function of $Imp(t)$ and $\varepsilon_{t}$ is the residue. Let $\hat{s_{t}}$ be an estimate of $g(t)$, using weighted least squares, where the weight for $(t, Imp(t))$ is large if $t$ is close to $t_k$ ($k=1,2,3...n$ and $n$ is the number of points) and small if it is not. Now we give the precise details of the smoothing procedure. We shall construct a weighted function $M$, which decreases for increasing nonnegative $t$, used to calculate weights $m_k(t)$ for each $t_k$. The weight function is generally set as the cubic function, that is defined as,
\begin{equation}
    M(x)= 
    \begin{cases}
      \left(1-|x|^3\right)^3, & |x|<1 \\
      ~~~~0, & |x|\ge 1
    \end{cases}.
\end{equation}

Setting the fraction coefficient as $f\in(0,1]$, we round $fn$ to get the data width $r =\left[ f n \right]$. We center the $M$ at $Imp(t)$ and scale it so that $M$ at the $r^{th}$ nearest neighbor of $Imp(t)$ first becomes zero. For each point, let $h_t$ be the distance from $t$ to the $r^{th}$ nearest neighbor of $Imp(t)$. For $k=1,2,3...n$, we set
\begin{equation}
    m_{k}(t)= M\left(h_{t}^{-1}(t_{k}-t)\right).
\end{equation}
For each point, the estimated parameters should minimize
\begin{equation}
    \sum_{k=1}^{n} m_k(t)\left[Imp(t_k)-g(t)\right]^2.
   \label{Eq:wls}
\end{equation}
Thus we get the fitted value $\hat{s_{t}}$ of the regression at $t$. In fact, these estimates are the coefficients for the $Imp(t_k)$ that arise from the regression.

Based on such a procedure, we remove the temporal trend of the time series.  The assumption of $\varepsilon_{t}$ allows us to construct a first-order autoregression model,
\begin{equation}
  \varepsilon_{t}=\phi\varepsilon_{t-1}+e_{t},
   \label{Eq:ar}
\end{equation}
where $e_{t}$ is the random disturbance term. Cook's distance can be calculated as follows,
\begin{equation}
d_{t}=\frac{\left[\hat{\Phi}-\hat{\Phi_{t}}\right]^{T}\Sigma\Sigma^{T}\left[\hat{\Phi}-\hat{\Phi_{t}}\right]}
{\hat{e_{t}}^{2}},
\end{equation}
where $\hat{\Phi}$ is the estimated parameter vector and $\hat{e_{t}}^{2}$ is the estimated random disturbance term of the AR(1) model in Eq.~(\ref{Eq:ar}), $\Sigma$ is the vector of explanatory variables, and $\hat{\Phi_{t}}$ is a new parameter vector containing the remaining sample data information except the data at period $t$.

To define shocks, we need first to determine the supply baseline over a certain time interval as the reference. We select the mean and median of food import trade volumes within a certain time interval $l$ in the past as the baseline, 
\begin{equation}
    Imp^{\mathrm{base}}(t) = {\mathrm{mean}}\left\{Imp(t-1), \cdots, Imp(t-l)\right\},
\end{equation}
or
\begin{equation}
    Imp^{\mathrm{base}}(t)= {\mathrm{median}}\left\{Imp(t-1), \cdots, Imp(t-l)\right\},
\end{equation}
where $t>l$. The number of years used for shock identification is 
\begin{equation}
    N_{\mathrm{Identify}} =  N_{\mathrm{Origin}}-l.
\end{equation}
The downward import fluctuation magnitude $\Delta Imp(t)$ is calculated as follows,
\begin{equation}
 \Delta Imp(t)=Imp^{\mathrm{base}}(t)- Imp(t).
\end{equation}
A shock is identified at $t$ when Cook's distance $d_t$ is greater than a certain threshold $d_c$ and the import volume is smaller than the supply baseline $Imp^{\mathrm{base}}(t)$ in a certain time interval,
\begin{equation}
    Shock_{t}= 
    \begin{cases}
      1, & ~~ {\mathrm{if}}~\Delta Imp(t)>0~{\mathrm{and}}~d_t>d_c\\
      0, & ~~ {\mathrm{if}}~\Delta Imp(t)\le 0
    \end{cases},
\end{equation}
where $\Delta Imp(t)$ is the shock magnitude.
The time to recovery from a food import shock is defined by the point at which the import flow returns to the level of 5\% before the shock, i.e.,
\begin{equation}
    \tau^{\mathrm{re}} = \min_{\tau}\{\tau: Imp(t+\tau)\geq 0.95  Imp^{\mathrm{base}}(t)\}.
\end{equation}

\begin{figure}[h!]
    \centering
    \includegraphics[width=0.345\linewidth]{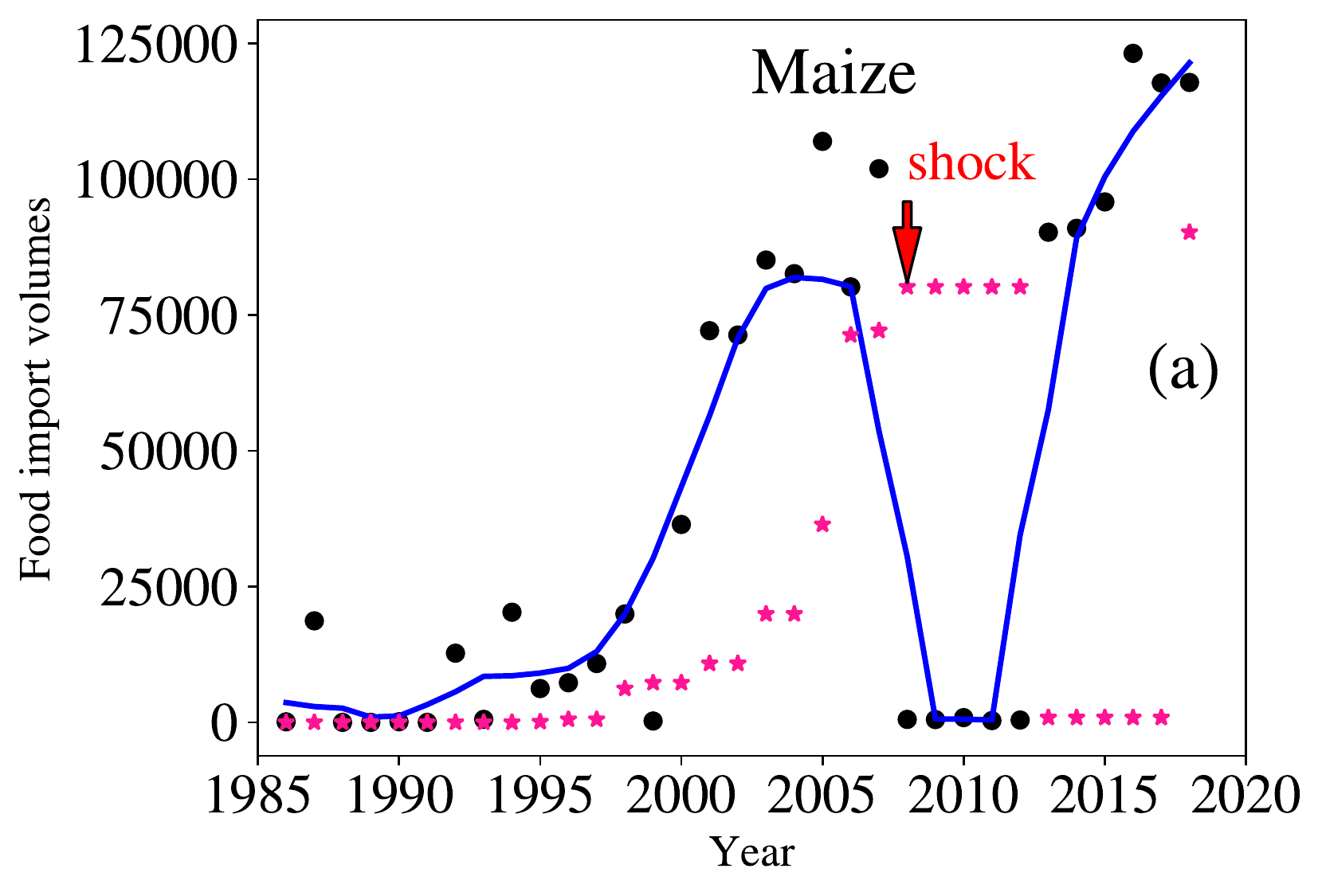}
    \includegraphics[width=0.32\linewidth]{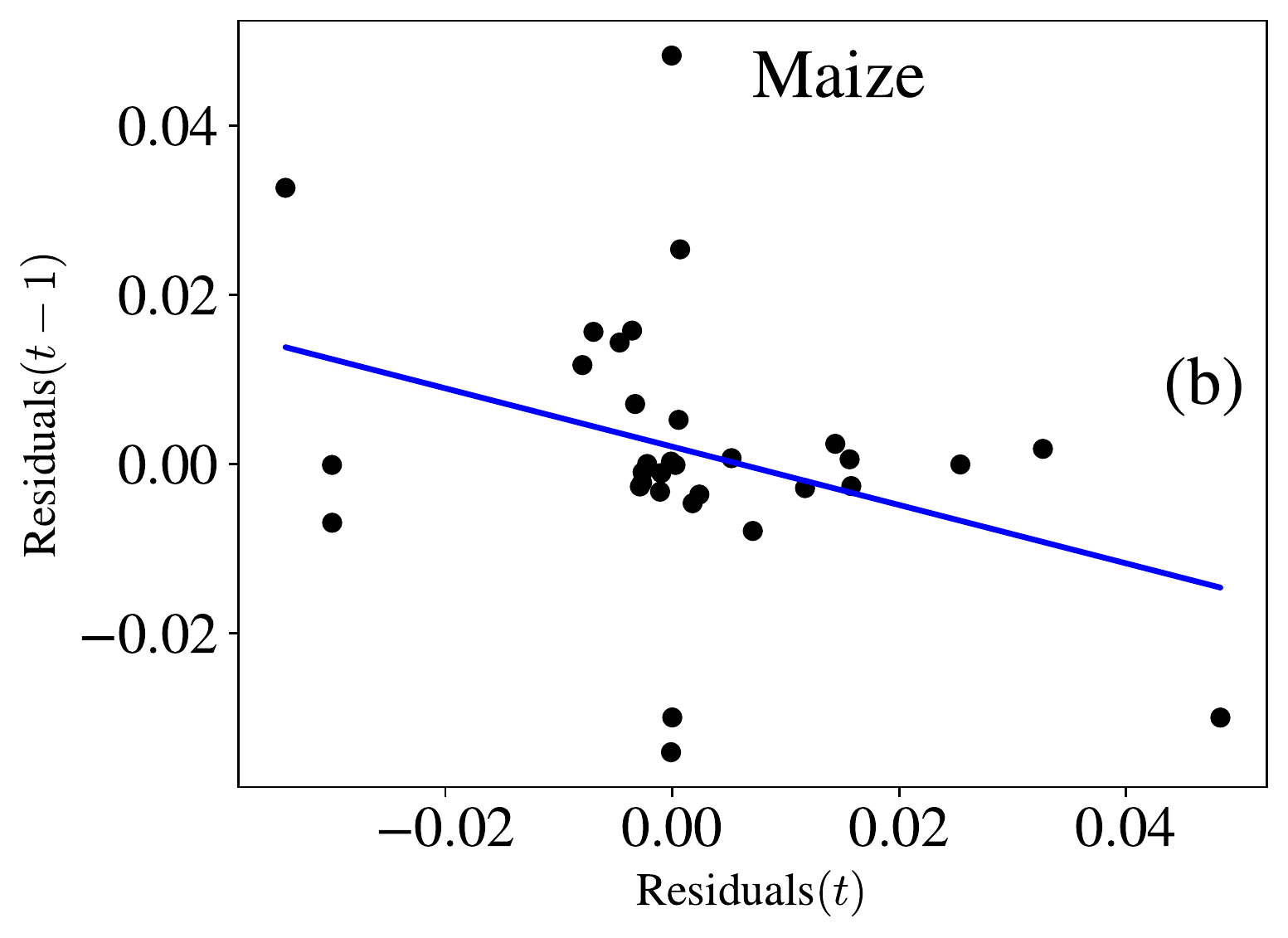}
    \includegraphics[width=0.305\linewidth]{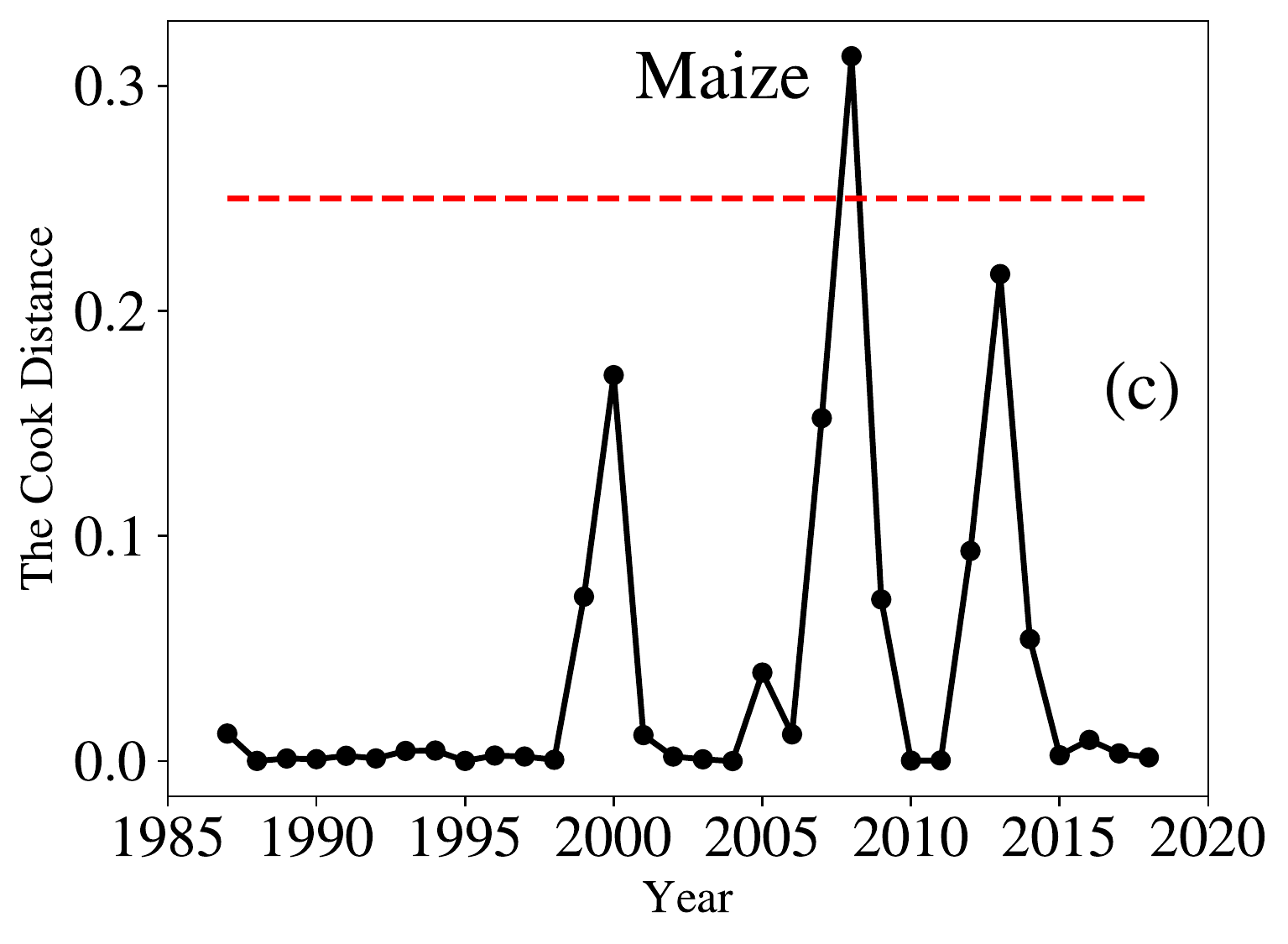}
    \caption{Import shock detection process for maize import of Eswatini (the second smallest country in continental Africa). (a) Locally weighted polynomial regression (LOWESS) model fitted to food trade volume time series. The black dots are maize import volumes and the blue line is the fitted curve. The pink dots are median values of food import volumes 9 years before shocks. (b) Regression of LOWESS model residuals $\epsilon_t$ against the lag-1 residuals $\epsilon_{t-1}$. (c) The point in 2008 is identified as an outlier from regression in (b) using Cook's distance. The maize import volume in 2008 is smaller than the supply baseline shown as pink dots in plot (a). It indicates that a maize import shock occurred in 2008. }
    \label{Fig:iCTN:Shock:Detection}
\end{figure}

Figure~\ref{Fig:iCTN:Shock:Detection} illustrates the identification process of the maize import shocks in Eswatini. Fig.~\ref{Fig:iCTN:Shock:Detection}(a) is the LOWESS regression result, Fig.~\ref{Fig:iCTN:Shock:Detection}(b) is the first-order autoregression process of residual error, and Fig.~\ref{Fig:iCTN:Shock:Detection}(c) is the identification results of maize import shocks. Points with high Cook's distances ($> 0.25$) and positive shock magnitudes are identified as shocks. It can be seen from Fig.~\ref{Fig:iCTN:Shock:Detection}(a) that LOWESS can fit the time trend of the international maize import volumes well, and the residual autoregression results in Fig.~\ref{Fig:iCTN:Shock:Detection}(b) shows that there are outliers. Fig.~\ref{Fig:iCTN:Shock:Detection}(c) shows that the point at 2008 is an outlier and the import volume in 2008 is smaller than the median of food import volumes within nine years before 2008. It indicates that a maize import shock occurred in 2008. The global food price crisis of 2007/2008 and the subsequent economic recession dealt a severe blow to food security. Maize prices rose 80\% in 2007 \cite{Anriquez-Daidone-Mane-2013-FoodPolicy}, and many consumers are turning to other crops to meet their food needs. This shows that our method can indeed identify food import shocks driven by some extreme events \cite{Harttgen-Klasen-Rischke-2016-FoodPolicy}.

\subsection{Sensitivity analyses}

The identification result of import shocks depends on the setting of parameters, including the fraction $f$ in the LOWESS model, the threshold $d_c$ of Cook's distance, the duration $l$ of the supply baseline, and the statistical method $m$ (i.e. the mean or median) of the base values. To obtain an optimal parameter combination, we use different parameter settings to identify food import shocks. We perform the shock detection analysis using a range of values for LOWESS fraction ($f=0.2\sim 0.8$, by a spacing of 0.1), duration used for import volume baseline average ($l=3$, 5, 7, and 9 years), and average type ($m=$mean or median). By changing the parameters, $7\times4\times2=56$ kinds of recognition results are obtained. We construct a simple linear regression model between the median of the numbers of shocks identified under all parameter combinations and the numbers of annual shocks identified under a certain parameter combination through time. We select the parameter combination that minimize the sum of squared residuals (SSR) with the median of this range through time \cite{Gephart-Deutsch-Pace-Truell-Seekell-2017-GlobEnvironChange}. In this paper, when the fraction of the LOWESS model is 0.3, the duration of supply baseline is 9 years, and the average type is median, the sum of residual squares reaches the minimum. For this parameter combination, the sum of squared residuals is 82.22, and the maximum of SSR is 948.41 which is obtained under the parameter combination ($f=0.2$, $l=3$ and $m=$median). The SSR values under other combinations all exceed 90. Therefore, this optimal parameter combination is used for shock identification. The maximum and minimum annual shock frequencies are used as confidence intervals.

\begin{figure}[h!]
    \centering
    \includegraphics[width=0.475\linewidth]{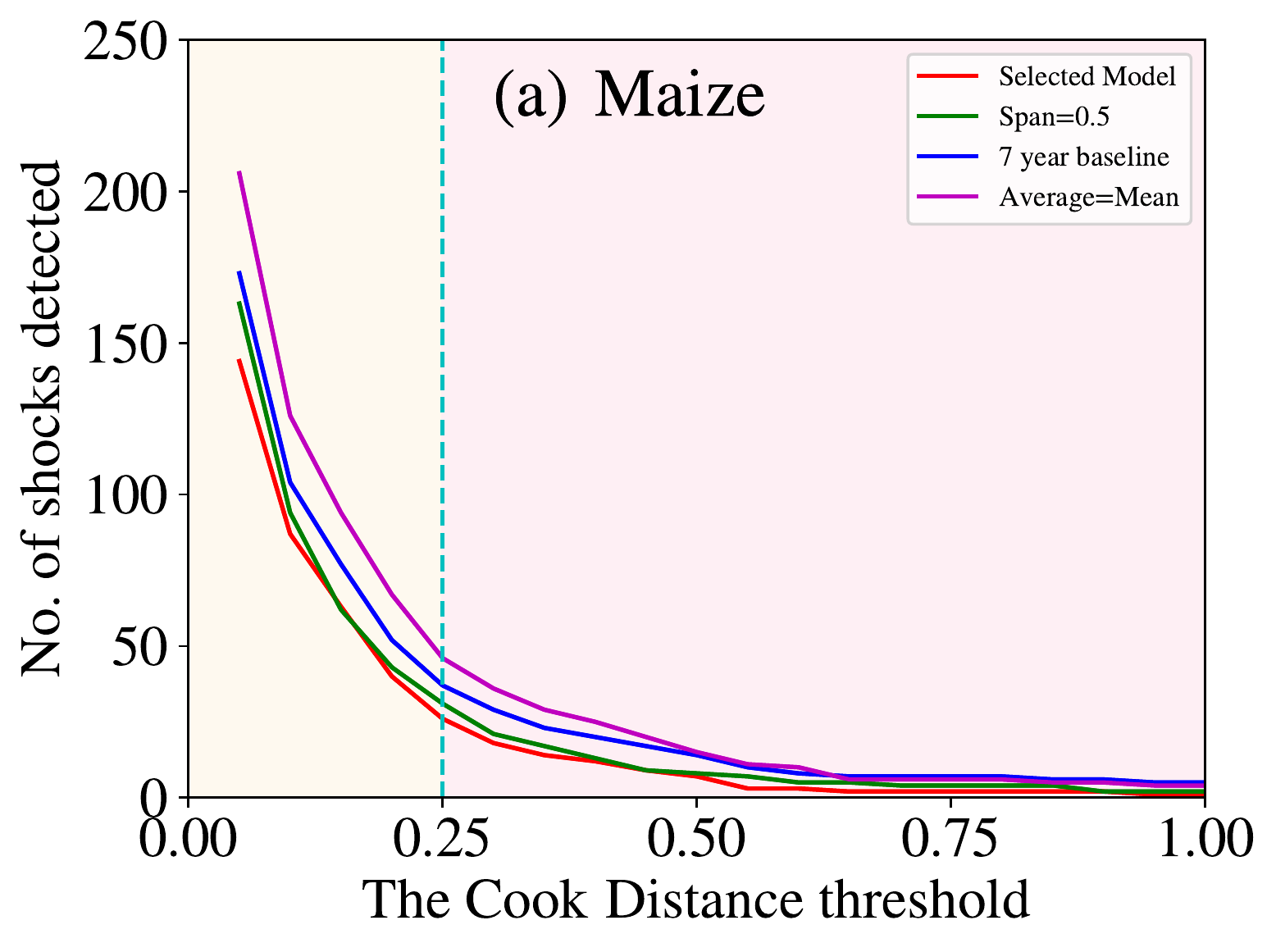}
    \includegraphics[width=0.475\linewidth]{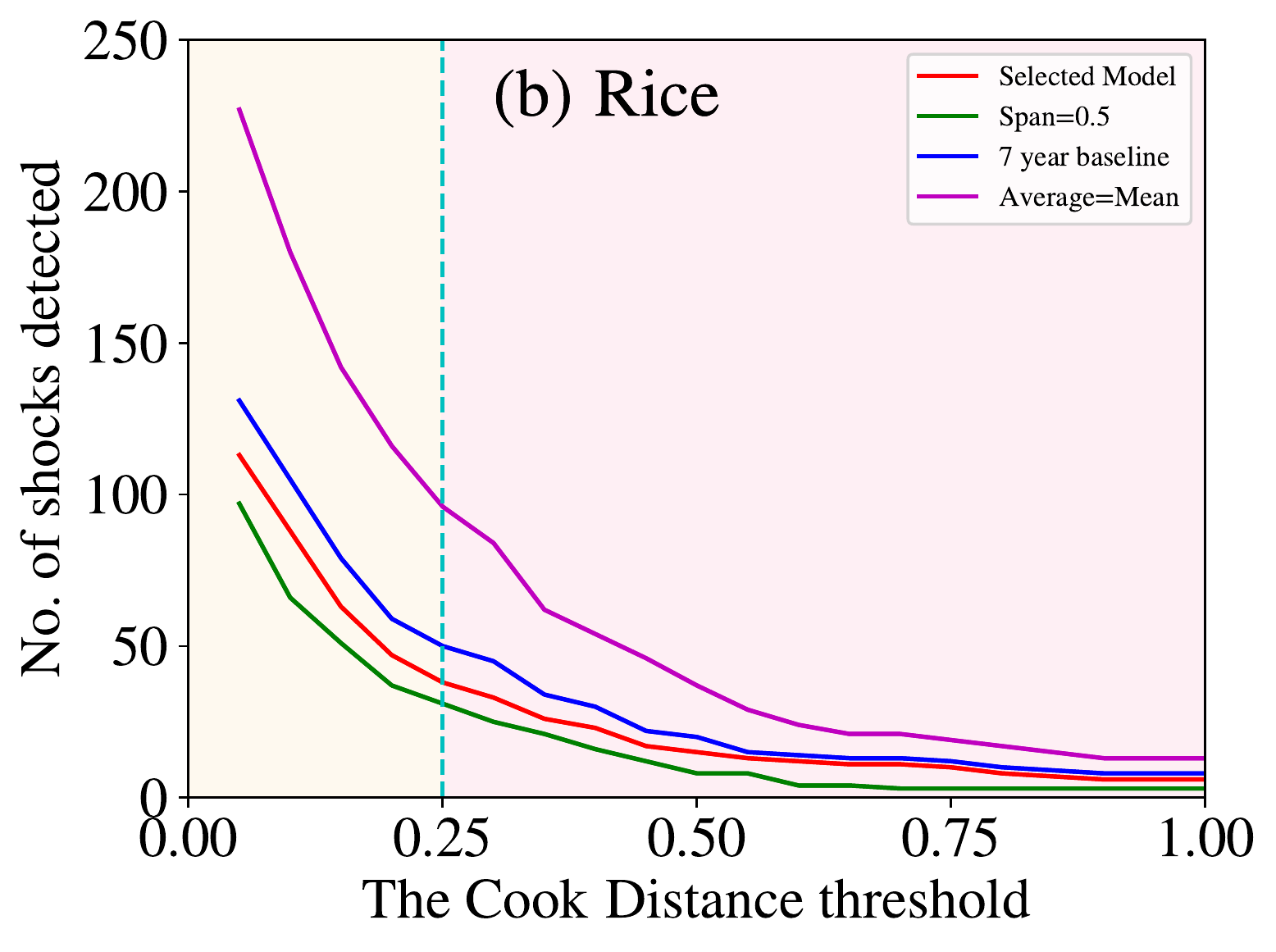}
    \includegraphics[width=0.475\linewidth]{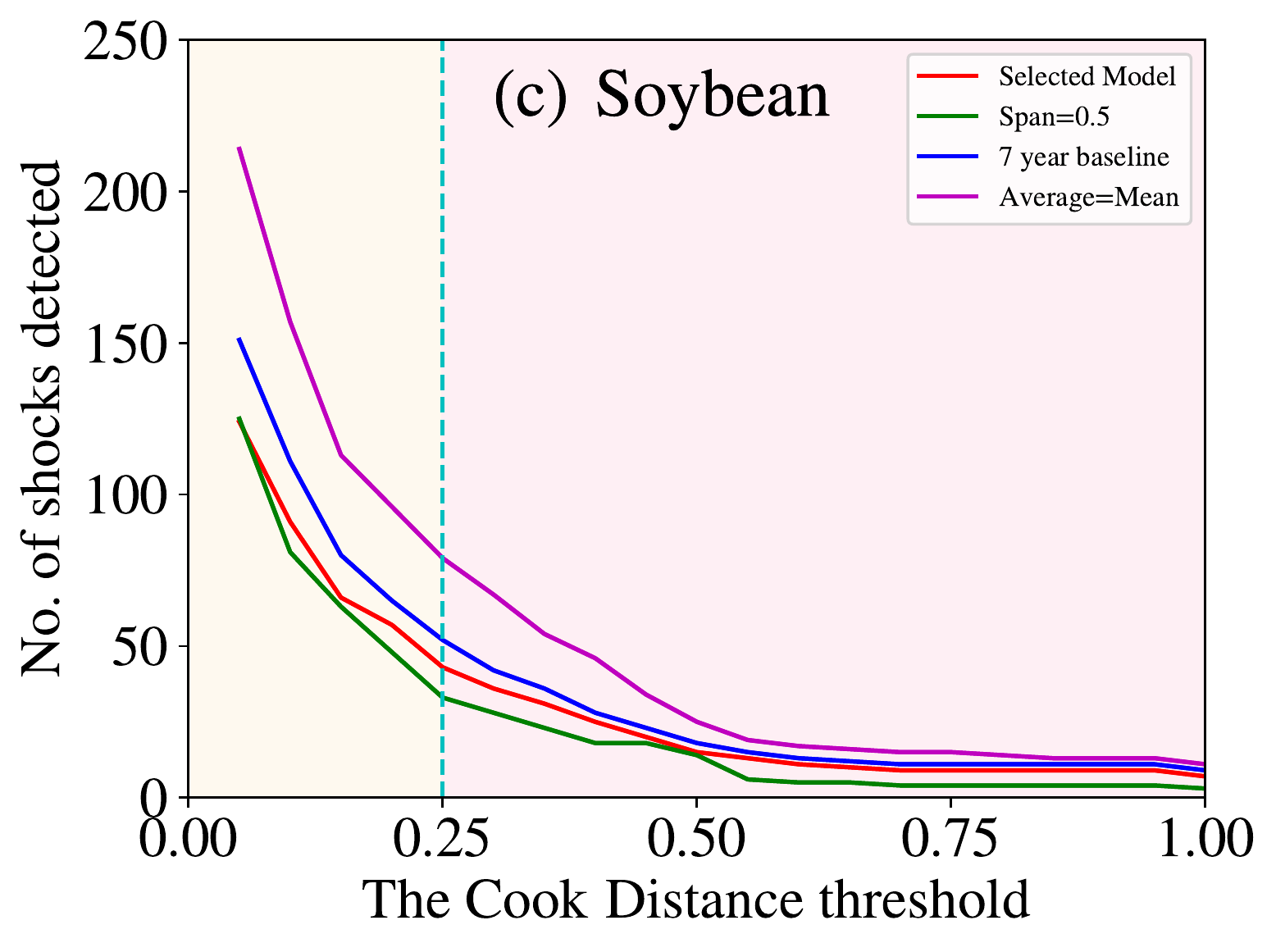}
    \includegraphics[width=0.475\linewidth]{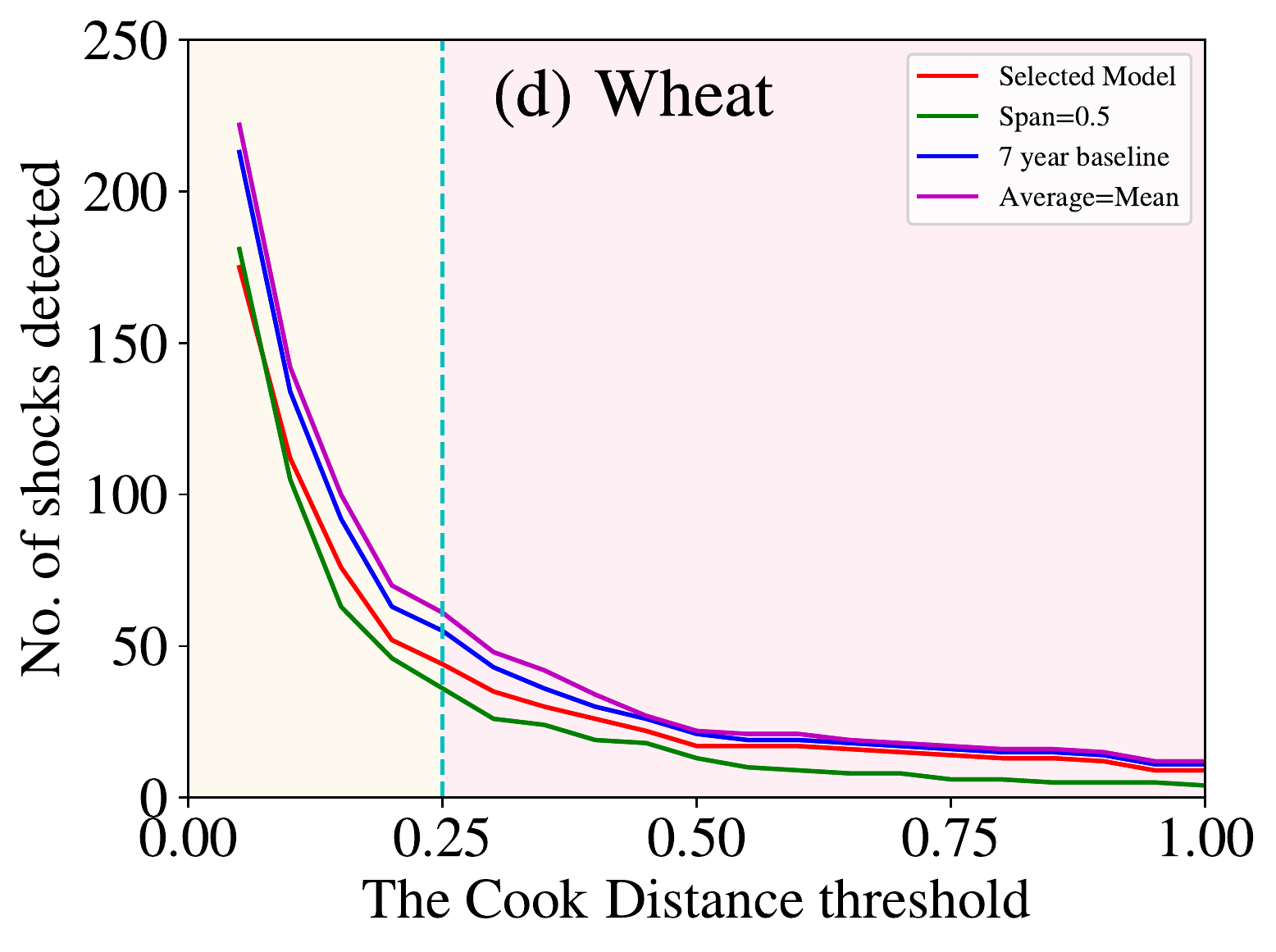}
    \caption{Comparisons of number of shocks detected in maize, rice, soybean and wheat time series with incremental changes to Cook's distance values. Lines represent either the combination of model parameters used in this study (`Selected Model', LOWESS span $f=0.6$, trade baseline $=7$ years and average type used = median), or repeated with changes to model span, trade baseline or average type. Vertical dashed line represents Cook's distance value of 0.25 used in this study.}
    \label{Fig:iCTN:CookDistance}
\end{figure}

To determine the threshold of Cook's distance for identifying outliers, we use different thresholds to calculate the shock frequencies. In previous studies, points with Cook's distance greater than 1 or $\frac{4}{n_t-p-1}$ ($n_t$ is the sample size of $\varepsilon_{t}$, and $p$ is the number of estimated parameters. Here, $n_t=32$ and $p=2$.) are generally considered as outliers \cite{Cottrell-Nash-Halpern-Remenyi-Corney-Fleming-Fulton-Hornborg-Johne-Watson-Blanchard-2019-NatSustain}. In this analysis, we integrate these two methods, set Cook's distance threshold $d_c$ varying from 0.05 to 1, and take into account the other three parameters to perform sensitivity analysis. Figure~\ref{Fig:iCTN:CookDistance} shows the comparisons of the number of shocks detected in the four food import time series with the change of Cook's distance threshold. We find that the value of 0.25 is the point in this relationship, reasonable across all crops, where the number of shocks detected begins to asymptote \cite{Cottrell-Nash-Halpern-Remenyi-Corney-Fleming-Fulton-Hornborg-Johne-Watson-Blanchard-2019-NatSustain}. The Cook's distance threshold selected in our study is close to that in other studies about identifying shocks \cite{Gephart-Deutsch-Pace-Truell-Seekell-2017-GlobEnvironChange,Cottrell-Nash-Halpern-Remenyi-Corney-Fleming-Fulton-Hornborg-Johne-Watson-Blanchard-2019-NatSustain}.

\subsection{Food import diversity and dependency ratio}

There are a number of factors that could illustrate the consequences of shocks over time or in particular regions. It is particularly hard to standardize the method of identifying food import shocks. Besides, it is difficult to consider all shock causes since the global food system is exposed to climatic, economic and geopolitical risks. Here, we just discuss some factors affecting the international food trade. Our data-driven approach can complement the understanding of food import trade shocks over time and across regions. 

Diversity is an important determinant of food system resilience. In the international food trade networks, some economies are more vulnerable to food production shortages abroad because of their high trade dependency, while import diversity can buffer food security against this risk \cite{Marchand-Carr-Dell'Angelo-Fader-Gephart-Kummu-Magliocca-Porkka-Puma-Ratajczak-Rulli-Seekell-Suweis-Tavoni-D'Odorico-2016-EnvironResLett}. There are many ways to measure diversity. We learn from the method adopted by Gomez et al. \cite{Gomez-Mejia-Ruddell-Rushforth-2021-Nature} to calculate the diversity of food supply and use Shannon Entropy to calculate the food import diversity of an economy,
\begin{equation}
  D_{i}^{food}(t) = -\frac{1}{k^{\mathrm{in}}_{i,food}(t)}
  \sum_{e_{ji}\in{\mathscr{E}}}\frac{w_{ji,food}(t)}{s^{\mathrm{in}}_{i,food}(t)}\mathrm{log}\frac{w_{ji,food}(t)}{s^{\mathrm{in}}_{i,food}(t)},
  \label{Eq:Diversity}
\end{equation}
where $D_{i}^{}(t)$ is the import diversity of the food of economy $i$ in year $t$, $k^{\mathrm{in}}_{i,food}(t)$ is the in-degree of economy $i$, ${\mathscr{E}}=\{e_{ij}\}$ is the collection of network edge $e_{ij}$, and $w_{ji,food}(t)$ means the trade volumes of the food flowing from economy $j$ to economy $i$ in year $t$.

Some economies are not self-sufficient in domestic food production and need to import food from other economies to meet their domestic food demands. Food import dependency ratio (IDR) can be used to measure percentage of an economy's dependency on imports of food to meet domestic needs. The higher is the value of import dependency ratio, the more food supply needs to be imported, namely,
\begin{equation}
 IDR_{i}^{}(t) = \frac{s^{\mathrm{in}}_{i,food}(t)}{P_{i,food}(t)+s^{\mathrm{in}}_{i,food}(t)-s^{\mathrm{out}}_{i,food}(t)},
  \label{Eq:IDR}
\end{equation}
where $P_{i,food}(t)$ is the production of food in year $t$.

\section{Identification of import shocks}
\label{S3:Results}

\subsection{Patterns and trends of shocks}

As illustrated in Table~\ref{Tab:iCTN:shock:numbers}, there are 206 importers of maize, 213 importers of rice, 199 importers of soybean, and 201 importers of wheat. In total, we have 819 time series of length 33. For each time series, we perform the shock identification procedure. Since $l=9$ and $t>l$, shocks prior to 1995 cannot be identified. Hence, we have $N_{\mathrm{Identify}}=33-9=24$. During 24 years from 1995 to 2018, we identify 33 maize import shocks, 43 rice import shocks, 42 soybean import shocks, and 52 wheat import shocks.

\begin{table}[!ht]
    \caption{The numbers of time series (or importing economies), observations and shocks identified for maize, rice, soybean and wheat during the period 1986--2018. We use $l=9$ to determine the baseline so that the shocks are identified between 1995 and 2018.}
    \smallskip
    \setlength\tabcolsep{7.5mm}
    \centering
    \begin{tabular}{cccccc}
    \toprule
        Crops & Number of time series &  Number of observations &  Number of shocks \\
    \midrule
        Maize   &  206  &  6798  & 33 \\ 
        Rice    &  213  &  7029  & 43 \\ 
        Soybean &  199  &  6567  & 42 \\ 
        Wheat   &  201  &  6633  & 52 \\ 
    \midrule
        Total  &   819  &  27027 & 170 \\ 
    \bottomrule
    \end{tabular}
\label{Tab:iCTN:shock:numbers}
\end{table}

\begin{figure}[h!]
    \centering
    \includegraphics[width=0.475\linewidth]{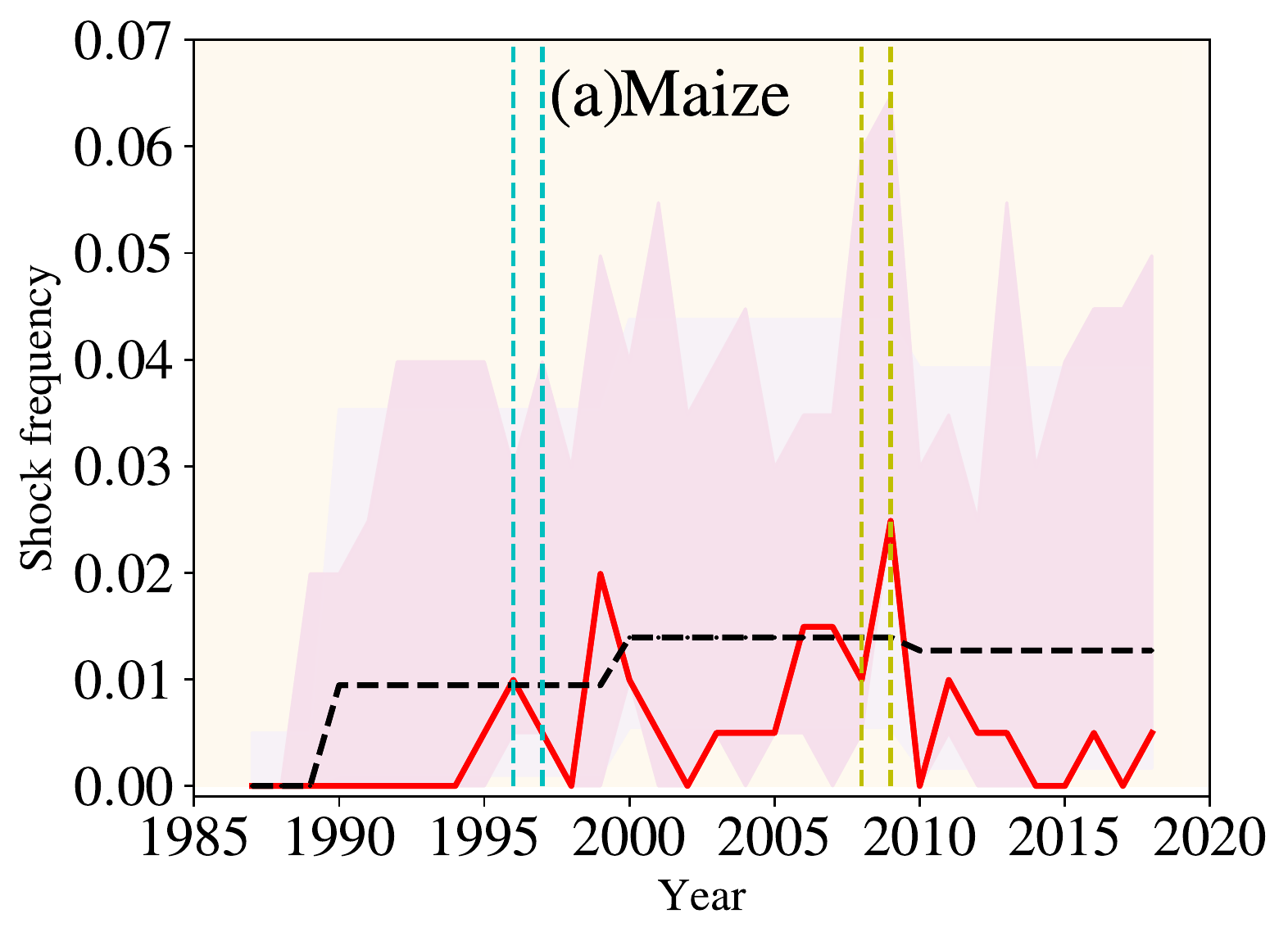}
    \includegraphics[width=0.475\linewidth]{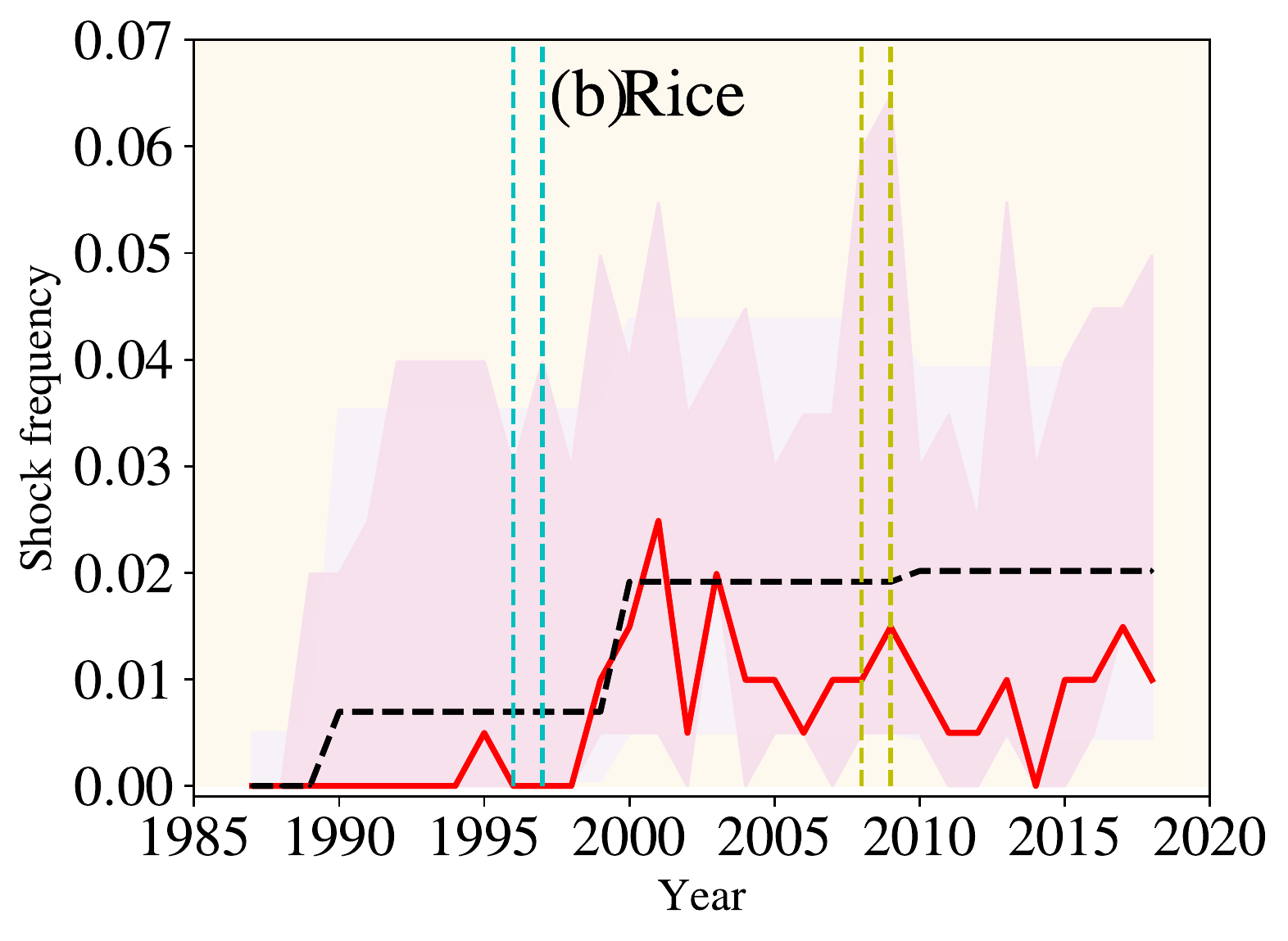}
    \includegraphics[width=0.475\linewidth]{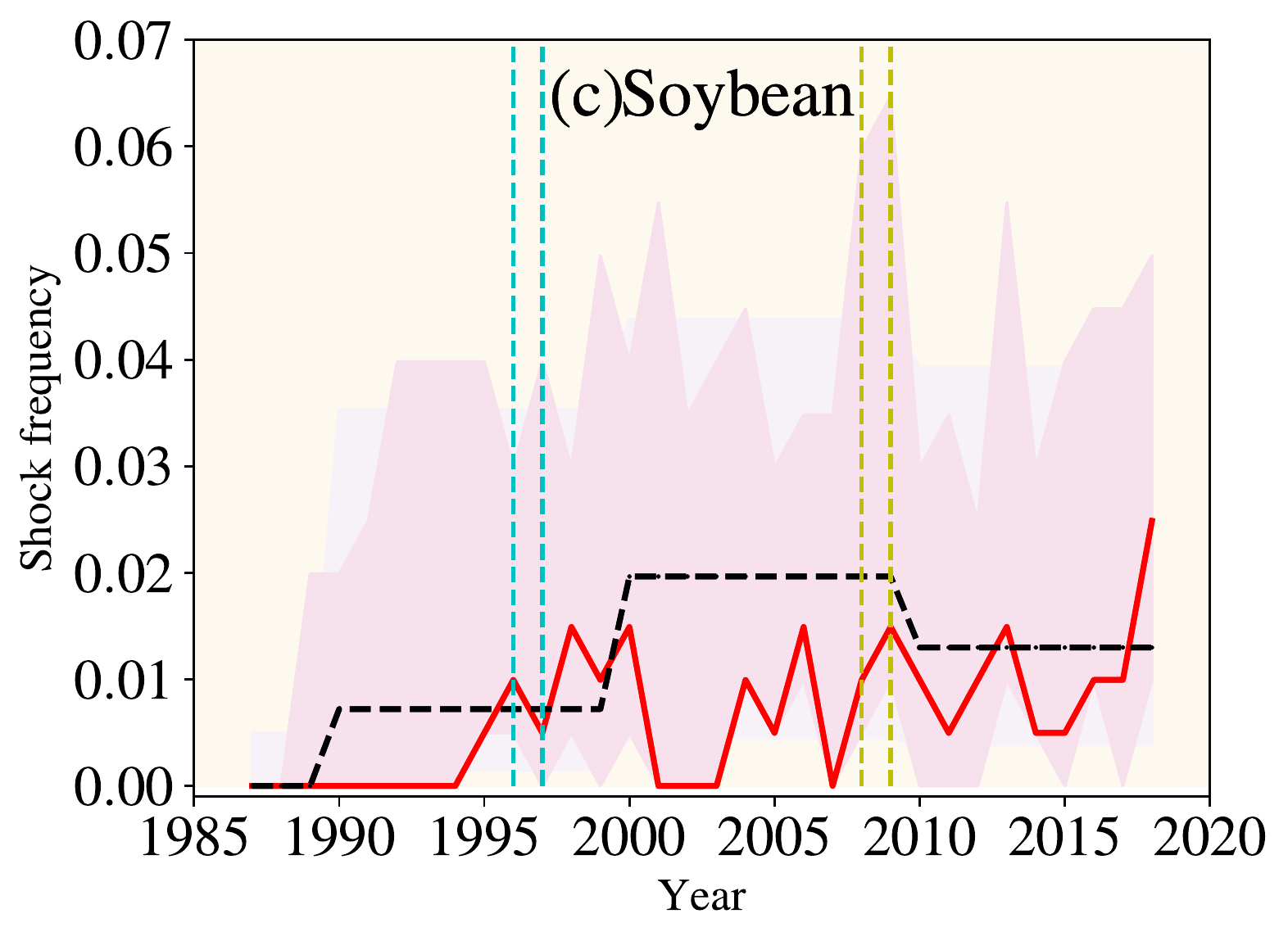}
    \includegraphics[width=0.475\linewidth]{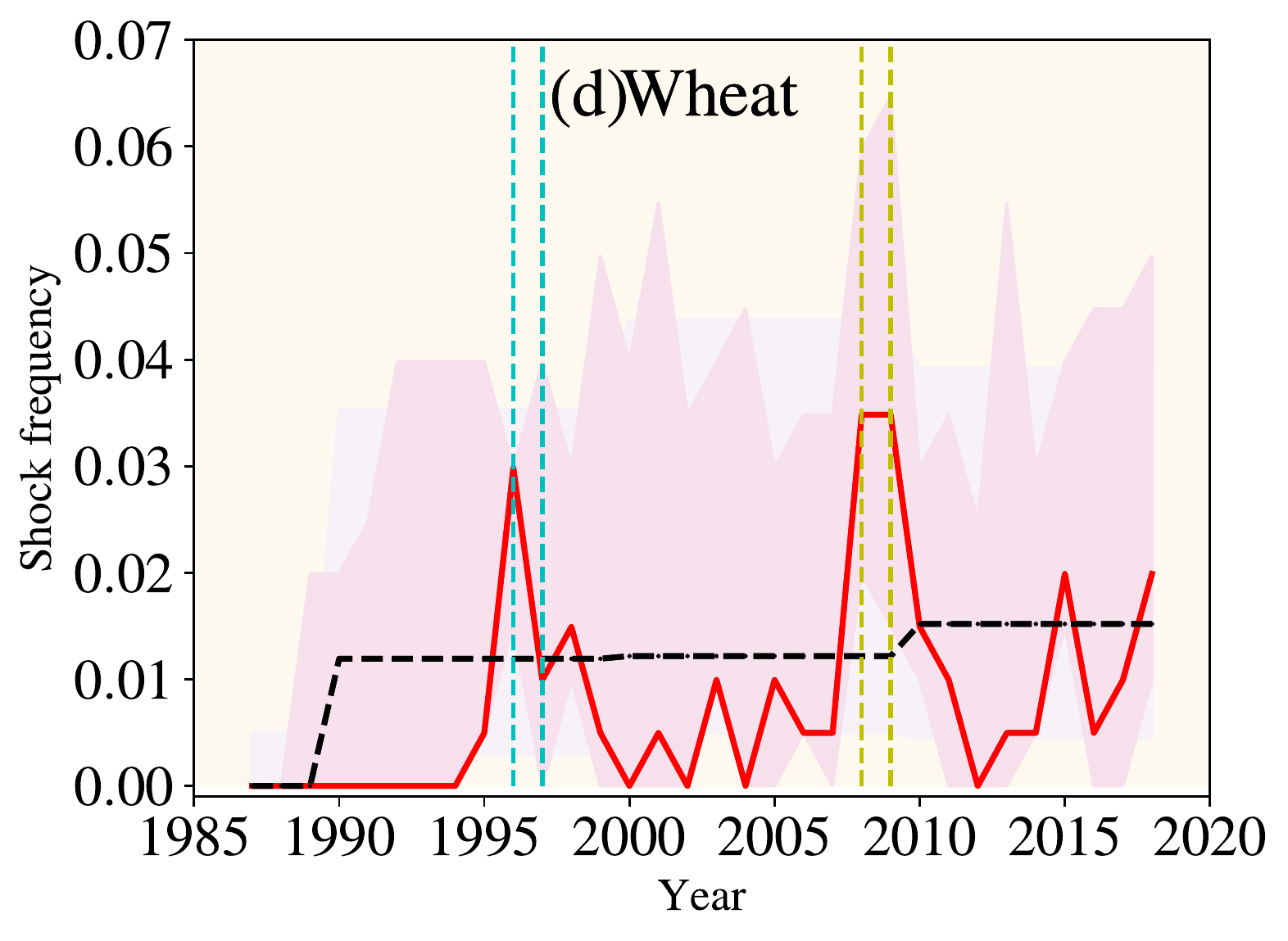}
    \caption{Trends in food import shock frequency in maize (a), rice (b), soybean (c), and wheat (d) from 1987-2018. The red lines in the time series indicate the annual shock frequency from the shocks identified in this study. The shock frequency of a given food is equal to the proportion of the number of economies experiencing shocks to the number of economies participating in food import trade in that year. The light pink confidence interval describes the plausible range of frequencies under different combinations of LOWESS model span ($f=0.2\sim0.8$), supply baseline duration ($l=3$, 5, 7 or 9 years), and types of averaging used for the baseline (mean or median). The dashed black line is the decadal mean of the red line. The dark pink band is the decadal minimum and maximum of the confidence interval. The events corresponding to the dotted blue and yellow lines are the financial crisis (1996/1997) and the food price crisis (2008/2009).}
    \label{Fig:iCTN:ShockFreq:4Crops}
\end{figure}

For a given crop in each year, we denote $\mathcal{N}_{{\mathrm{shock}},t}^{crop}$ as the set of economies with shocks identified in the import of $crop$. We calculate the import shock frequency for each crop, which is equal to the proportion of the number of economies experiencing at least a shock to the number of economies participating in the import trade  in that year \cite{Tu-Suweis-D'Odorico-2019-NatSustain}
\begin{equation}
    f_{\mathrm{shock},t}^{crop} = \frac{ \#(\mathcal{N}^{crop}_{{\mathrm{shock}},t})}{\#({\mathcal{N}}_{t}^{crop})},
\end{equation}
where ${\mathcal{N}}^{crop}_{t}$ is the set of economies importing $crop$ in a given year $t$, $\#(\cdot)$ is the counting function.
Figure~\ref{Fig:iCTN:ShockFreq:4Crops} shows the trend of the shock frequency from 1995 to 2018. We obtain a range of annual shock frequencies calculated by changing parameters mentioned in the shock detection method in Section~\ref{S2:ImportShock:DetectionMethod}. The red lines show the annual shock frequency under the optimal parameter combination with $f=0.3$, $l=9$, $d_c=0.25$ and $m=$median. The light pink confidence interval presents the range of frequencies under different parameter combinations. The dashed black line is calculated by averaging the shock frequencies during the decades (1995--2000, 2001--2010, and 2011--2018). It is equal to the decadal mean of the red line. The dark pink band describes the decadal minimum and maximum of the confidence interval. It shows that the shock frequencies of the four crops fluctuated considerably over time, yet were much higher in some years, especially during 1995--2000 and 2005--2010. This might be related to the economic crisis of 1997/1998 \cite{Rosegrant-Ringler-2000-FoodPolicy} and the food price crisis of 2007/2008 \cite{Ahmed-2014-AfrAsianStud,Headey-2011-FoodPolicy}. In 2008, as food prices soared, global food insecurity became severe.

The shock frequencies of different crops have different characteristics. For maize, significant food import shocks occurred in 1996, 1999, 2000, 2006--2009 and 2011, especially in 1999 and 2009. There were significant shocks in the rice import in 1999--2001, 2004, 2006, 2008--2010, 2012--2013 and 2016--2018, especially in 2001 and 2003. For soybean, there were significant shocks from 1996 to 2000, 2003 to 2006, and 2008 to 2018, especially in 2018. For wheat, there were significant shocks in 1996--1998, 2003, 2005, 2007--2011, 2015 and 2017--2018, especially in 1996 and 2008--2009. The trends of decadal average shock frequency present that rice and soybean have higher shock frequencies during 2000--2010, while rice and wheat have higher shock frequencies during 2010--2018. It indicates that different crops have unique international trade patterns. It is thus unreasonable to simply aggregate all kinds of crops to a total international food trade network when considering food security.

\begin{figure}[h!]
    \subfigbottomskip=-1pt
    \subfigcapskip=-5pt
    \centering
     \subfigure[]{\label{level.sub.1}\includegraphics[width=0.475\linewidth]{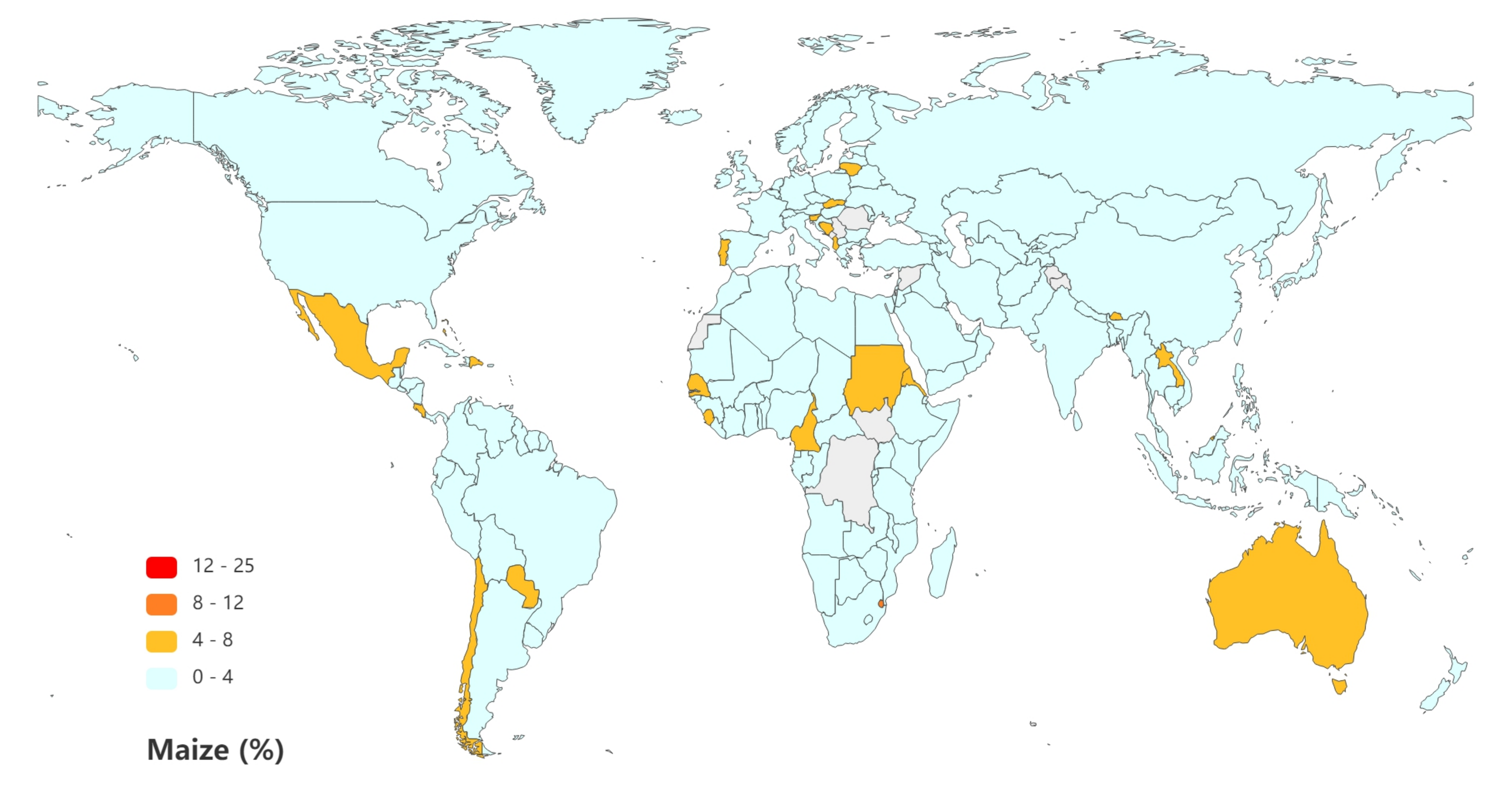}}
     \subfigure[]{\label{level.sub.2}\includegraphics[width=0.475\linewidth]{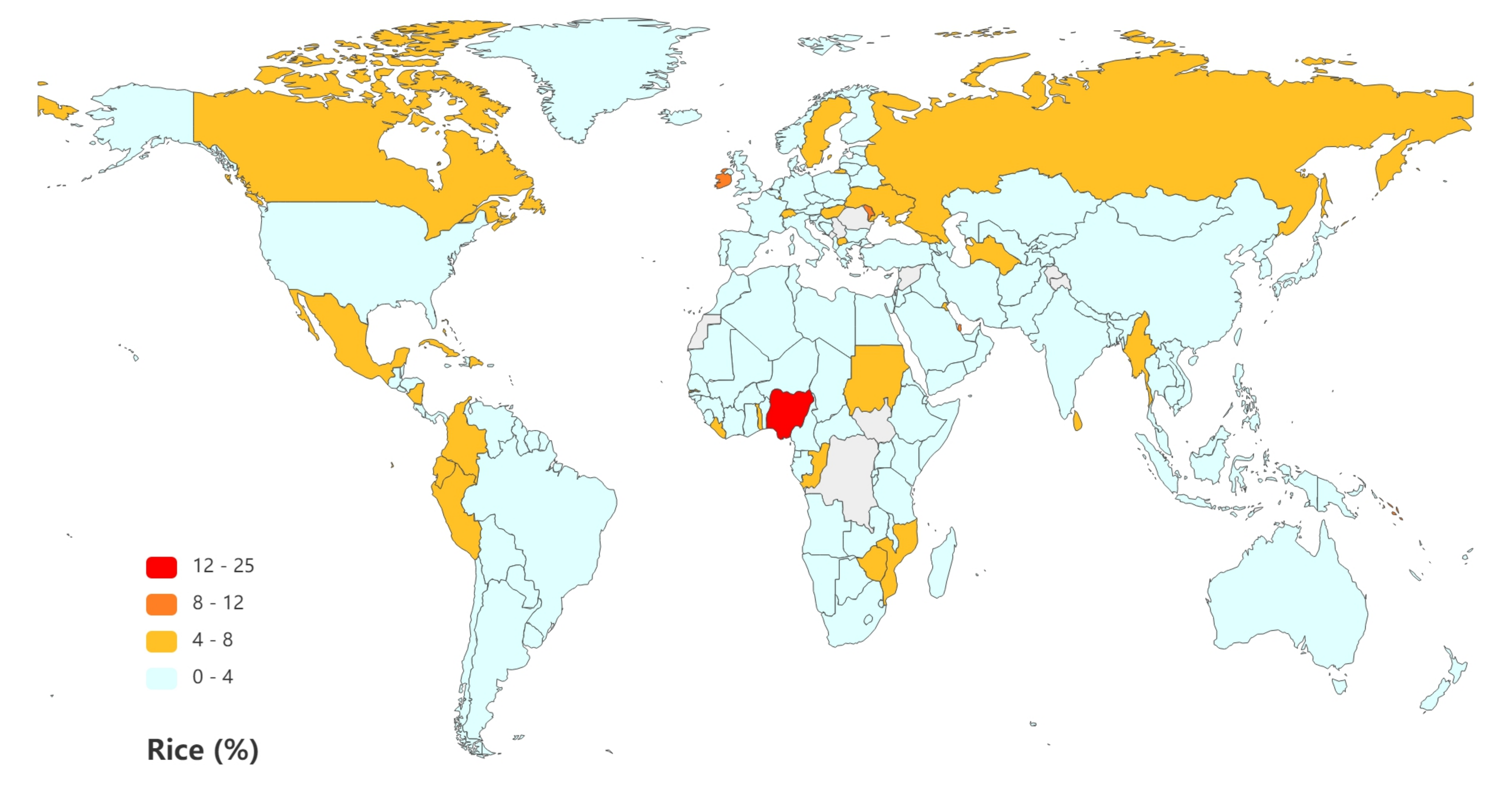}}
     \subfigure[]{\label{level.sub.3}\includegraphics[width=0.475\linewidth]{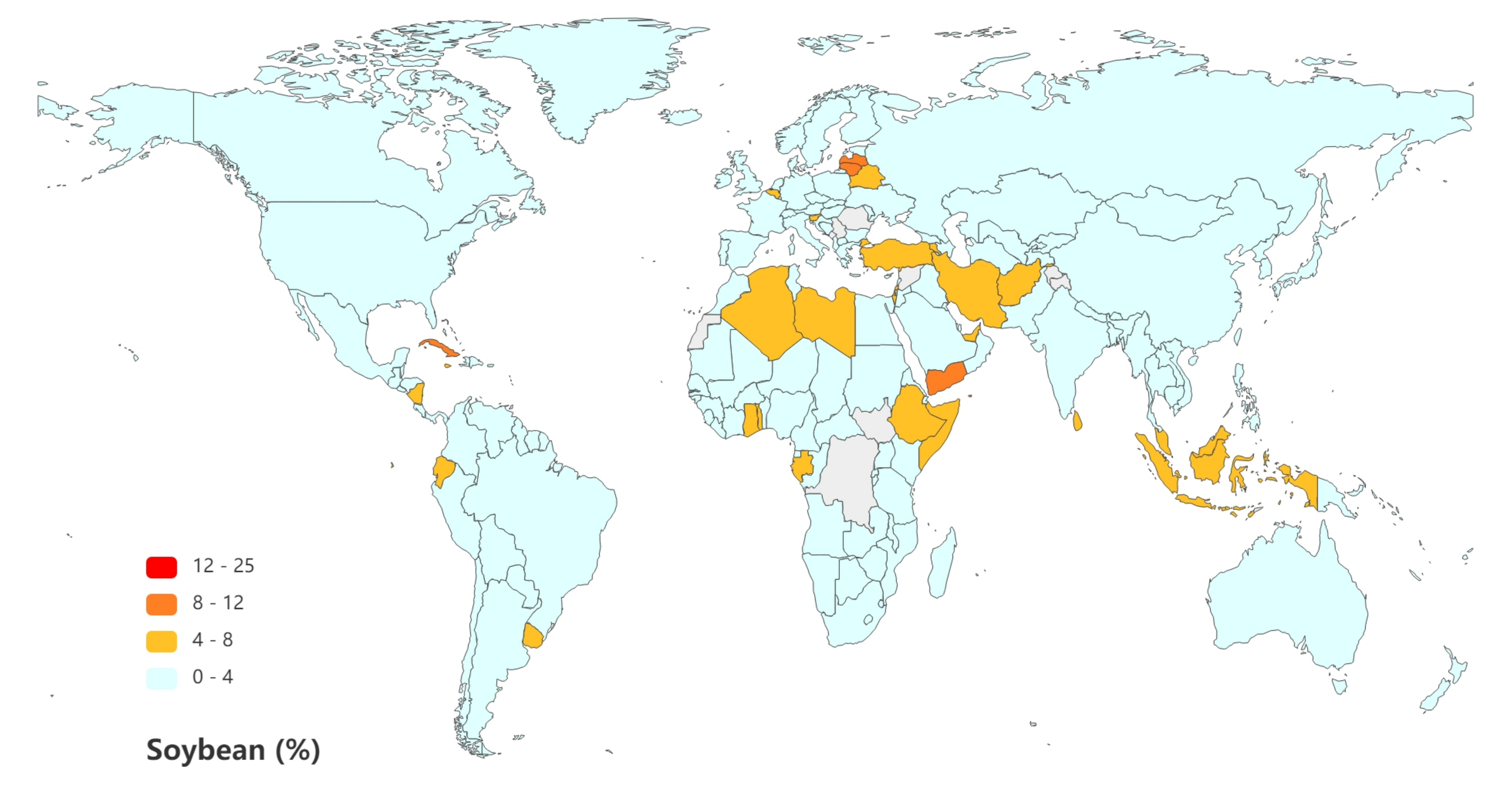}}
     \subfigure[]{\label{level.sub.4}\includegraphics[width=0.475\linewidth]{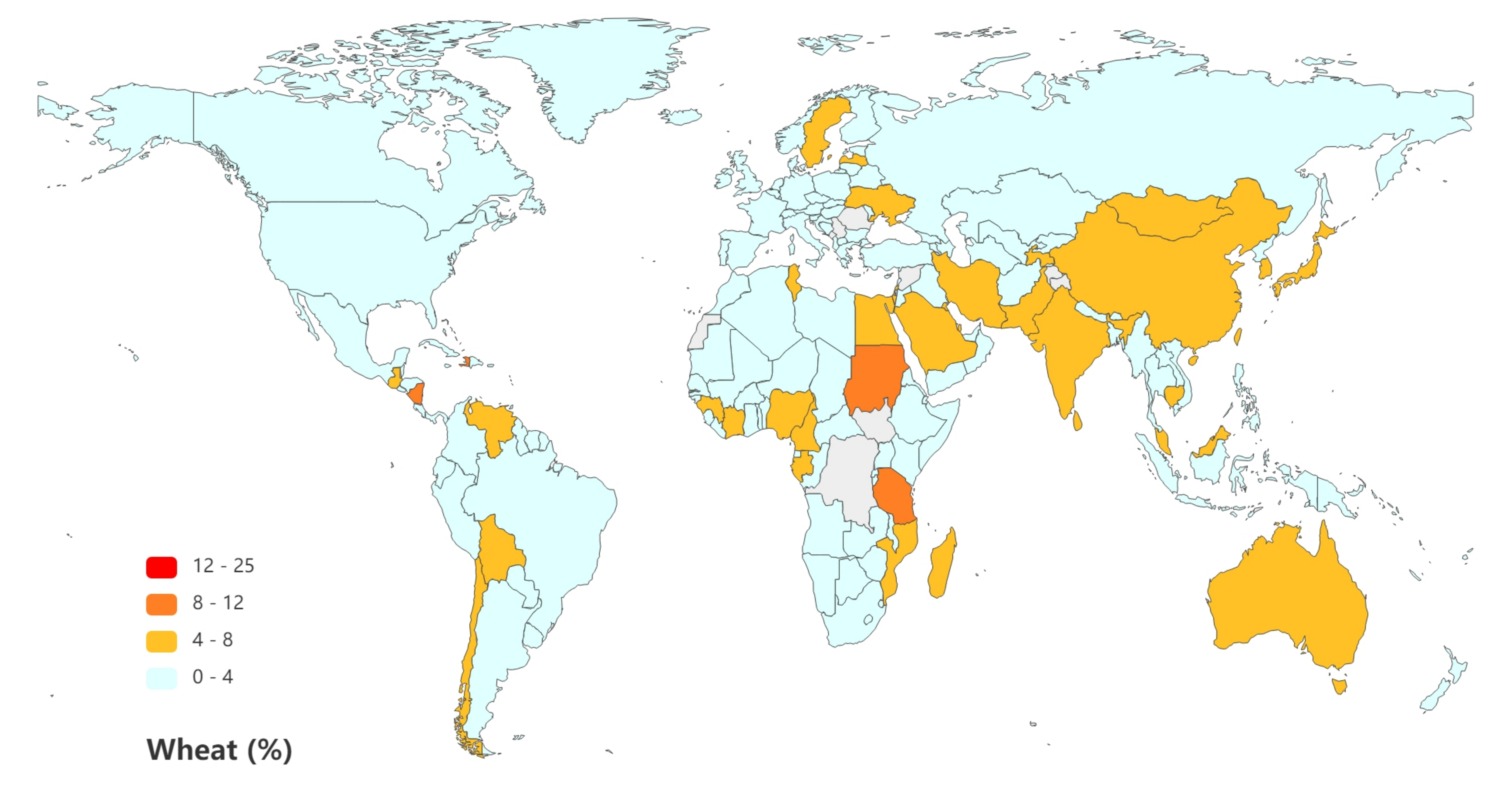}}
    \caption{Spatial distributions of economies' food import shock rates in maize (a), rice (b), soybean (c), wheat (d) sectors from 1995 to 2018. The shock rate of an economy refers to the number of shocks divided by the length of import time series. The darker the color, the more shocks happened. It is obvious that the shock frequencies of rice and wheat are relatively higher and regional distributions are wider.}
    \label{Fig:iCTN:shock:map}
\end{figure}

Different economies experienced different numbers of shocks. We define the shock rate of an economy as the number of shocks ($N_{{\mathrm{shock}},i}^{crop}$) divided by the number of observations of its import time series (here it is $N_{\mathrm{Identify}}=24$):
\begin{equation}
    r_i^{crop} = \frac{N_{{\mathrm{shock}},i}^{crop}}{N_{\mathrm{Identify}}}
\end{equation}
Figure~\ref{Fig:iCTN:shock:map} shows the spatial distributions of the import shock rates of the four crops from 1995 to 2018 at the economy level, where the darker the color is, the more shocks occur. 

It is obvious that rice and wheat import shocks covered wider regions. As can be seen from Fig.~\ref{Fig:iCTN:shock:map}(a), maize import shocks mainly concentrated in Africa and North America. Eswatini located in South Africa and Montserrat in North America had two maize import shocks. Australia, in Oceania, had a maize import shock in 1996. For the rice import, shocks appeared widely, which occurred in Europe, Africa, Asia and North America. Among them, Nigeria in Africa had the highest rice import shock rate, Ireland and Moldova in Europe, Qatar in Asia and Solomon Islands in Oceania had higher rice import shock rates. The soybean import shocks appeared more frequently in Asia and Europe. Five economies (Cuba, Latvia, Lithuania, Seychelles and Yemen) have experienced respectively two soybean import shocks in 24 years. The wheat import shocks have also hit a wide range of regions, mainly in Asia, Oceania, Africa and North America. Guadeloupe in North America had a higher wheat import shock rate. China had a wheat import shock in 1996. It is consistent with the event that China adjusted the wheat import policy in 1995. According to previous research, China was the world's largest wheat importer in 1995/96, but fell to the fifth in 1996/97 \cite{Carter-Zhong-1999-AmJAgrEcon}. From the above results, we preliminarily understand that shock rates varied across regions and crops. At the same time, shocks may be also related to the food trade policy of a given economy.

\begin{figure}[h!]
    \centering
    \includegraphics[width=0.475\linewidth]{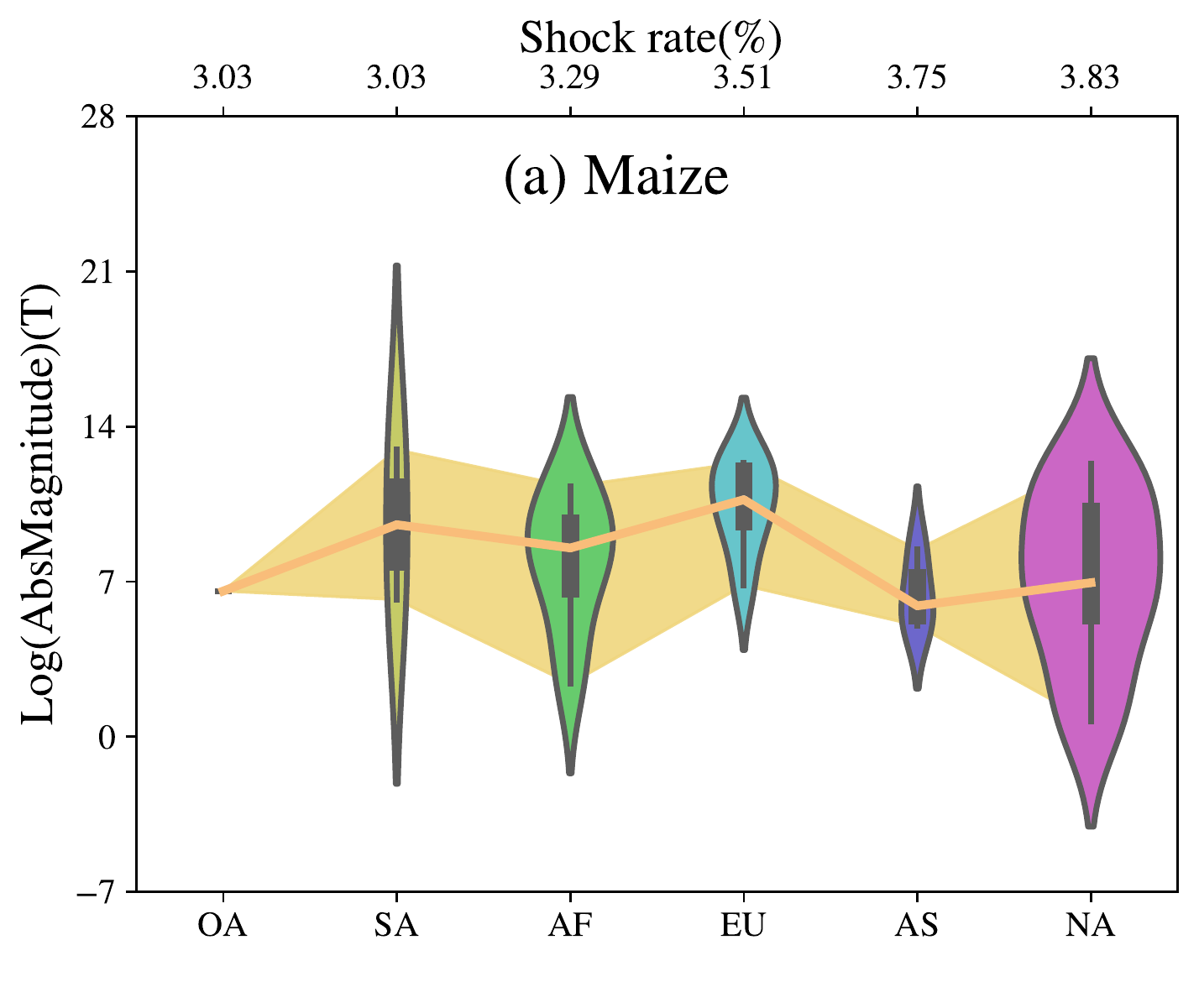}
    \includegraphics[width=0.475\linewidth]{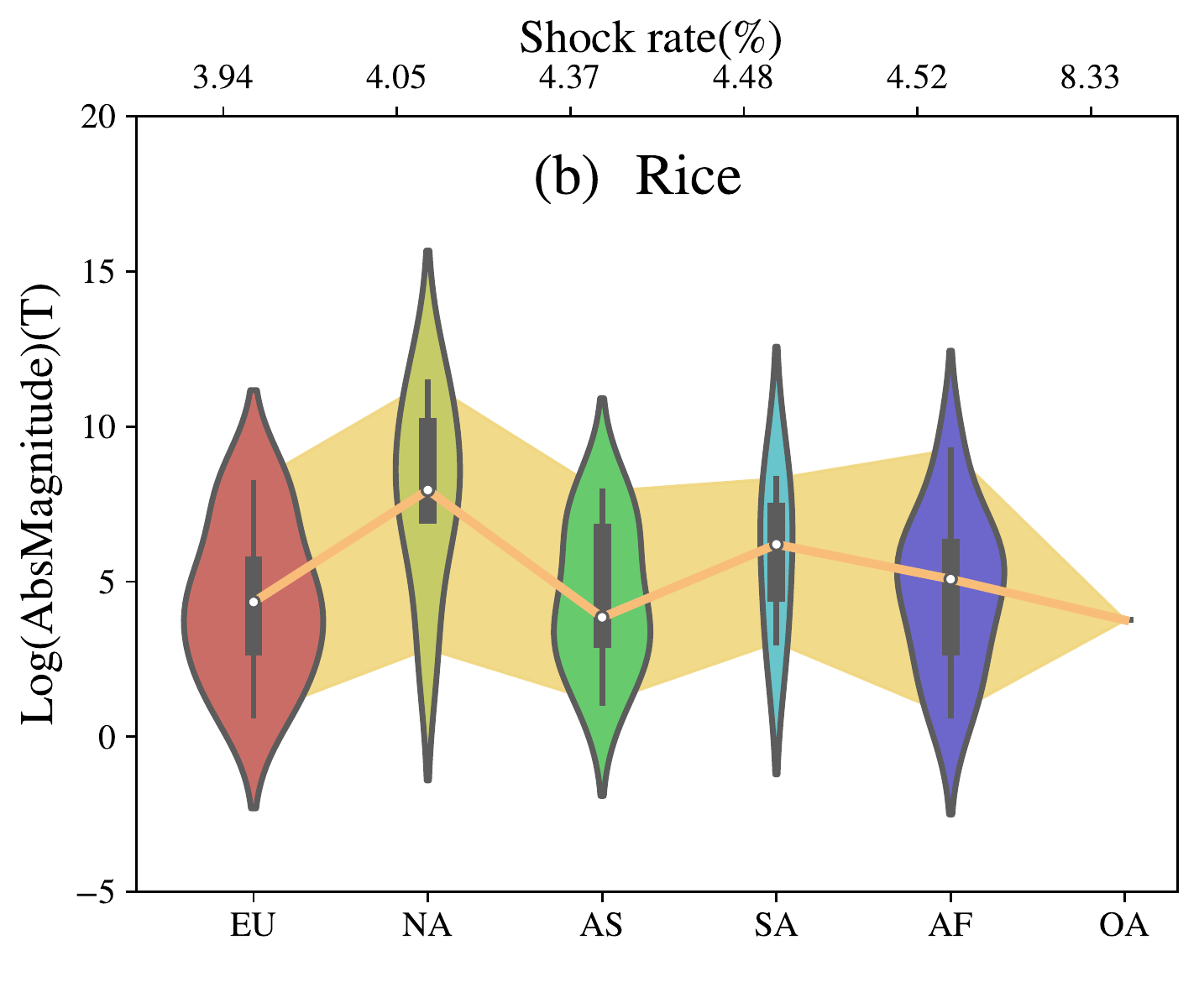}
    \includegraphics[width=0.475\linewidth]{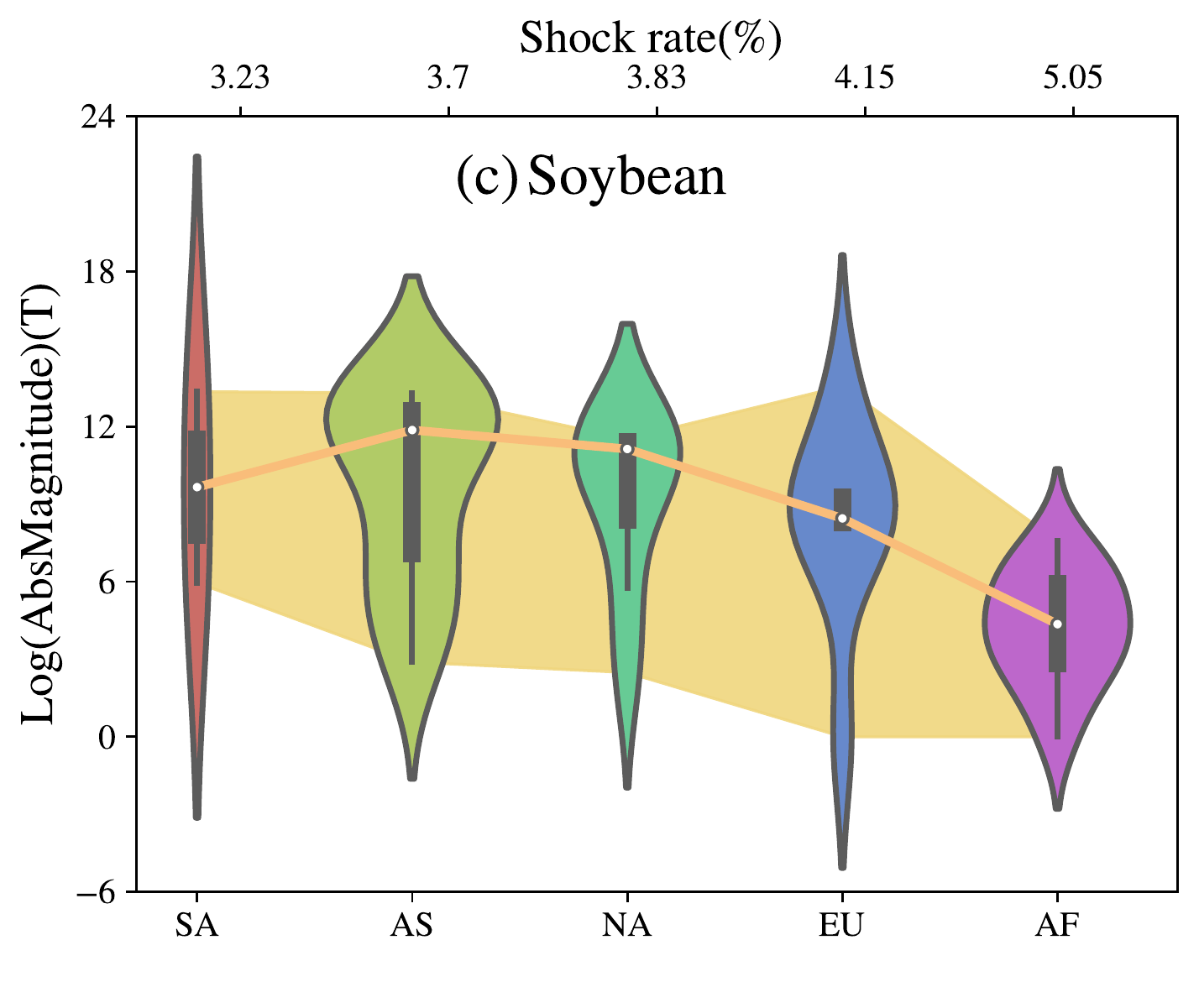}
    \includegraphics[width=0.475\linewidth]{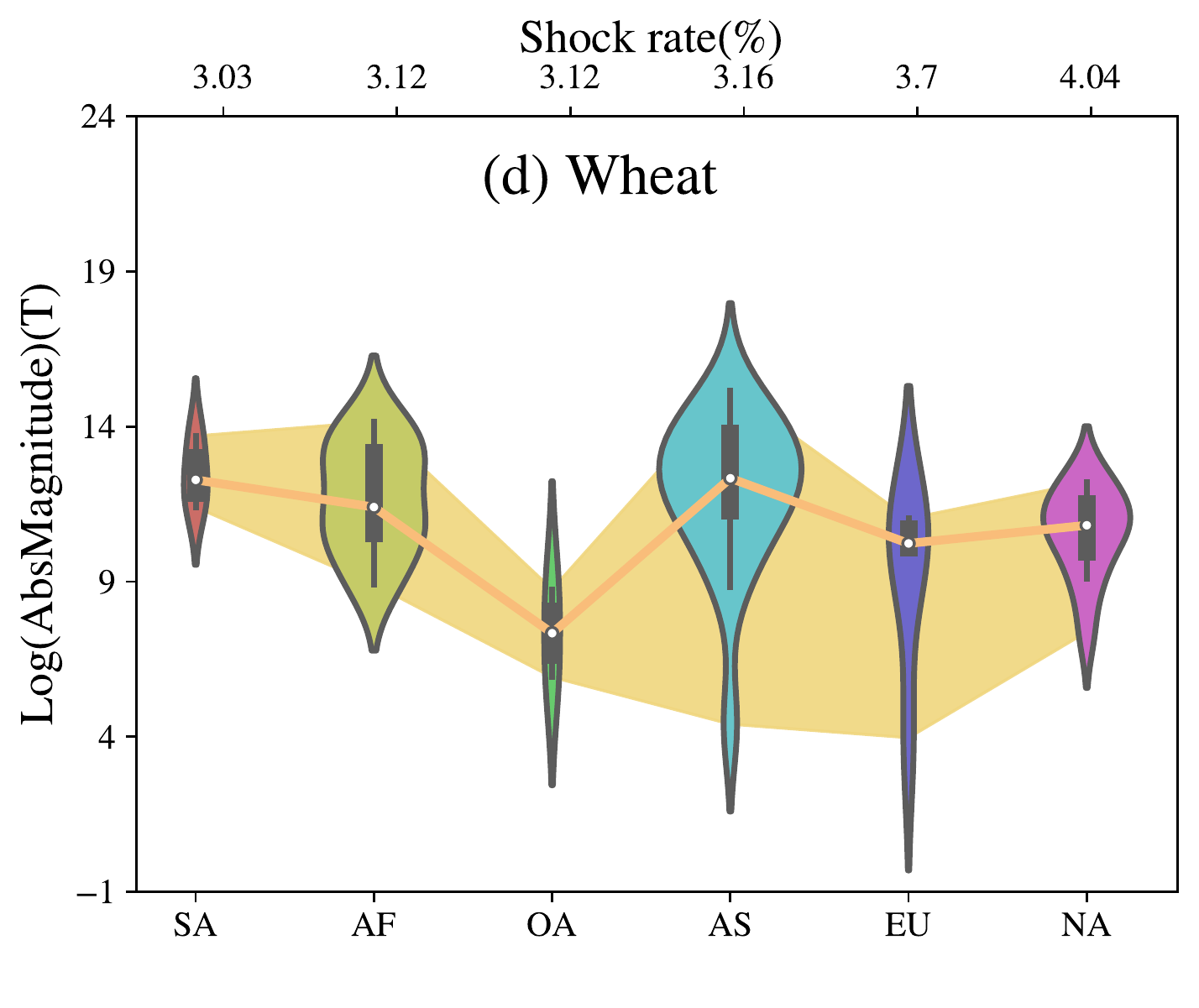}
    \caption{Shock rate $r_{\mathcal{C}}^{crop}$ and magnitude $\Delta s$ at the continent level for maize (a), rice (b), soybean (c) and wheat (d). The dark orange line shows the medians of the logarithmic shock magnitudes. The light orange shows the minimums and maximums. Note that AF refers to Africa, AS refers to Asia, EU refers to Europe, OC refers to Oceania, NA refers to North America, SA refers to South America, and AN refers to Antarctica. We sort the regions according to shock rates.}
    \label{Fig:iCTN:shock:magnitude}
\end{figure}

Patterns of shocks differed across regions. We calculate the shock rate of a given continent, which is defined as the ratio of the number of shocks to the number of observations in the continent from 1995 to 2018. 
\begin{equation}
    r_{\mathcal{C}}^{crop} = \frac{\sum_{i\in{\mathcal{C}}^{crop}}N_{{\mathrm{shock}},i}^{crop}}  {N_{\mathrm{Identify}}\times \#({\mathcal{C}}^{crop})},
\end{equation}
where ${\mathcal{C}}^{crop}$ is the set of economies in continent ${\mathcal{C}}$ importing $crop$. At the same time, we analyze the distributions of annual shock magnitudes $\Delta s$ in different continents from 1995 to 2018. In Figure~\ref{Fig:iCTN:shock:magnitude}, the lower abscissa shows the abbreviation of continents, the upper abscissa shows the shock rates of corresponding continents, and the ordinate shows the distributions of annual shock magnitudes. In general, the shock rates were not large, and the distributions of magnitudes were asymmetric. 

From Fig.~\ref{Fig:iCTN:shock:magnitude}(a), it can be seen that North America had a higher maize shock rate and most shocks were close to the medians of the shock magnitude. For maize import, the highest mean shock magnitudes occurred in Europe and South America, and the distribution of shock magnitudes in South America was obviously discrete. Only a few economies in Oceania import maize, including Australia, so the shock rates were low and the shock magnitudes have a very narrow distribution. As can be seen from Fig.~\ref{Fig:iCTN:shock:magnitude}(b), Europe had the highest rice import shock rate, but the shock magnitude was the second lowest. It suggests that shocks in Europe were relatively frequent but not large. North America had the second largest rice import shock rate and the highest mean shock magnitude. Most shocks in Asia were smaller than the medians. According to Fig.~\ref{Fig:iCTN:shock:magnitude}(c), the soybean import shocks mainly occurred in North America, and the shock magnitude was relatively large. South America had a low soybean import shock rate and the maximum of its shock magnitude was much larger. It was caused by Uruguay that had a soybean import shock with large magnitude in 2009. In the case of wheat import trade, as shown in Fig.~\ref{Fig:iCTN:shock:magnitude}(d), the region with the highest shock rate was Asia, in which China, Japan, South Korea and some other economies experienced wheat import shocks. Although China is a major wheat producer, it is also a major importer \cite{Carter-Zhong-1999-AmJAgrEcon}. Domestic wheat consumption in Korea\footnote{South Korea wheat imports forecast higher (2021), available athttps://www.world-grain.com/articles/16078-south-korea-wheat-imports-forecast-higher} and Japan\footnote{Japan's imported wheat price to hit 14-yr high amid Ukraine crisis (2015), available at https://english.kyodonews.net/news/2022/03/332c228c5346-japans-imported-wheat-price-to-rise-17-amid-ukraine-crisis.html} is very dependent on imports. In 1996 and around 2008, affected by the world food crisis \cite{McMichael-2009-AgricHumanValues}, many areas in Asia suffered import shocks.

These results together provide important insights into the spatial features of the shock magnitude and the shock rate. Overall, the food import shock rate and the shock magnitude are significantly different across regions, and also show unique characteristics over different crops. Comparatively speaking, North America had the higher food import trade shock rate, while South America had the larger shock magnitude.

\subsection{Recovery of shocks}

Food systems have resilience in the fake of shocks \cite{Butler-Davila-Alders-Bourke-Crimp-McCarthy-McWilliam-Palo-Robins-Webb-vanWensveen-Sanderson-Walker-2021-EnvironSciPolicy}. Shocks caused by a reduction in crop production or a sudden rise in food prices can be recovered as soon as possible after the end of these external disturbances, while shocks caused by political problems may be difficult to recover in a tick. We assume that the import recovers from a shock when the crop import volume reach again 95\% of the five-year average import volume prior to the shock. We analyze the number of recovered and not recovered cases and the shock recovery time across regions.

\begin{table}[!ht]
    \setlength{\tabcolsep}{6pt}
    \caption{Number of recovered and not recovered cases by region for maize (a), rice (b), soybean (c) and wheat (d). Recovery was defined as returning to within 5\% of the previous 5-year average before the shock. Note that AF refers to Africa, AS refers to Asia, EU refers to Europe, OC refers to Oceania, NA refers to North America, SA refers to South America, and AN refers to Antarctica.}
    \smallskip
    \centering
    \begin{tabular}{ccccccccccccccccccccc}
    \toprule
             & \multicolumn{6}{c}{Recovered} && \multicolumn{6}{c}{Not recovered} \\
        \cline{2-7}\cline{9-14}
        Crop & OC &  AS &  SA  & AF & EU & NA && OC &  AS &  SA  & AF & EU & NA  \\
    \midrule
        Maize & 1 & 3 & 2 & 7 & 4 & 8 && 0 & 0 & 0 & 1 & 2 & 5 \\ 
        Rice & 2 & 7 & 2 & 8 & 9 & 5 && 0 & 2 & 1 & 2 & 4 & 1 \\ 
        Soybean & 0 & 9 & 0 & 6 & 3 & 2 && 0 & 4 & 2 & 5 & 5 & 6 \\ 
        Wheat & 1 & 13 & 2 & 12 & 4 & 6 && 1 & 5 & 1 & 1 & 1 & 5 \\ 
    \midrule
        Total  &  4 & 32 & 6 & 33 & 20 & 21 && 1 & 11 & 4 & 9 & 12 & 17  \\ 
    \bottomrule
    \end{tabular}
\label{Tab:iCTN:shock:recovery}
\end{table}

Table~\ref{Tab:iCTN:shock:recovery} presents the number of recovered or not recovered shocks of the four crops across different regions from 1995 to 2018. 
It can be seen that North America had the most maize import shocks with 13 occurrence, but 5 shocks have not recovered. Maize import shocks in Asia, South America and Oceania were much less and have all recovered. 
It is apparent that Europe had the largest number of rice import shocks with 13 occurrences and 4 shocks are not recovered. 
For soybean import shocks, most cases in Asia have recovered, while more than half of cases in South America, Europe and North America have not returned to the corresponding pre-shock levels. 
For wheat import, there were 18 shocks in Asia with 13 shocks recovered, 13 shocks in Africa with 12 shocks recovered, and 11 shocks in North America with 6 shocks recovered.

These results suggest that shock recoveries in different regions presented diverse features, and varied over different crops. In general, most crop import shocks in Asia have recovered. Besides, most import shocks for maize, rice and wheat except soybean have recovered during the sample period. It is consistent with the scene that European economies increased domestic production by developing the cropping technologies for soybeans \cite{Dima-2015-AgricAgricSciProc}.

 
\begin{figure}[h!]
    \centering
    \includegraphics[width=0.475\linewidth]{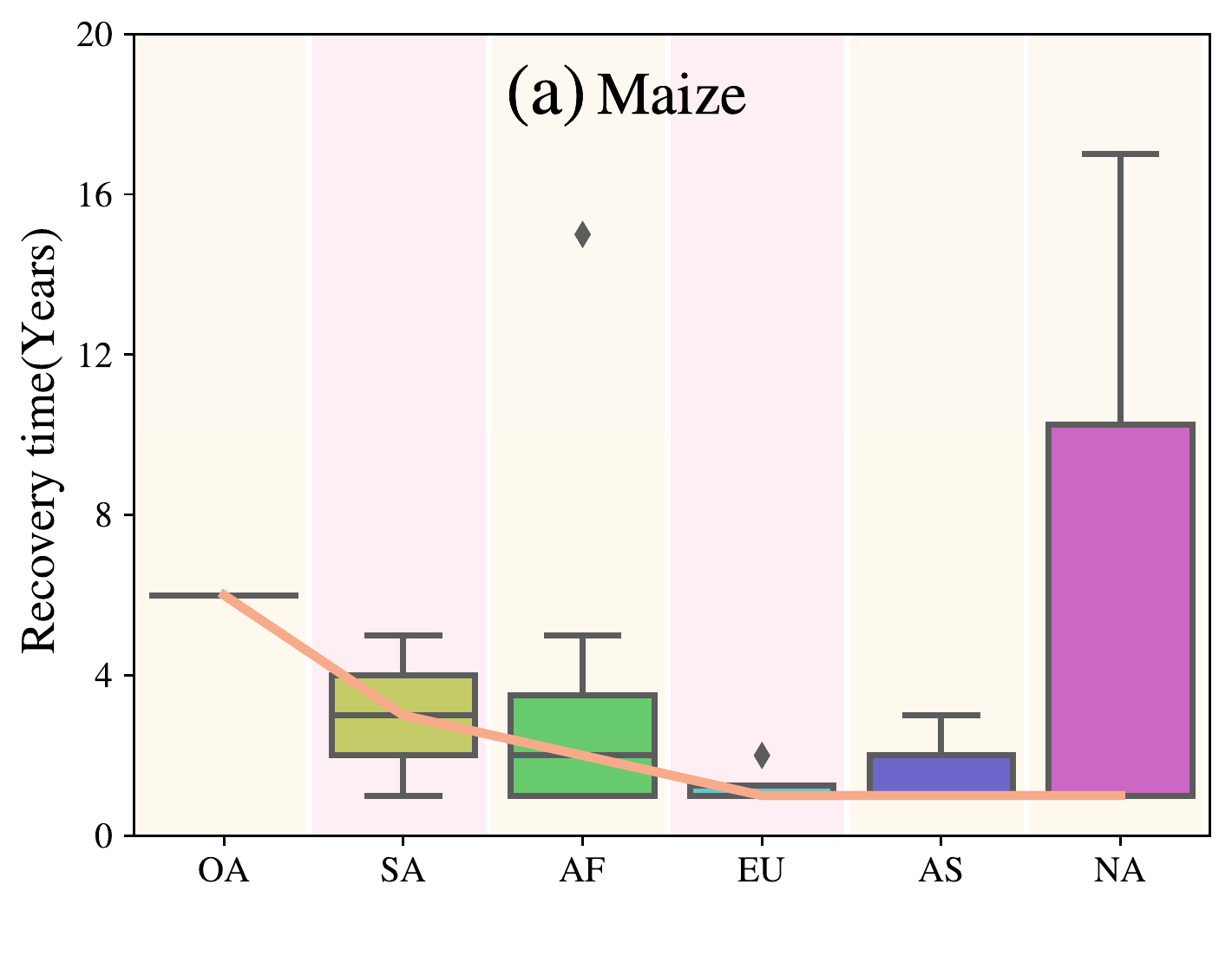}
    \includegraphics[width=0.475\linewidth]{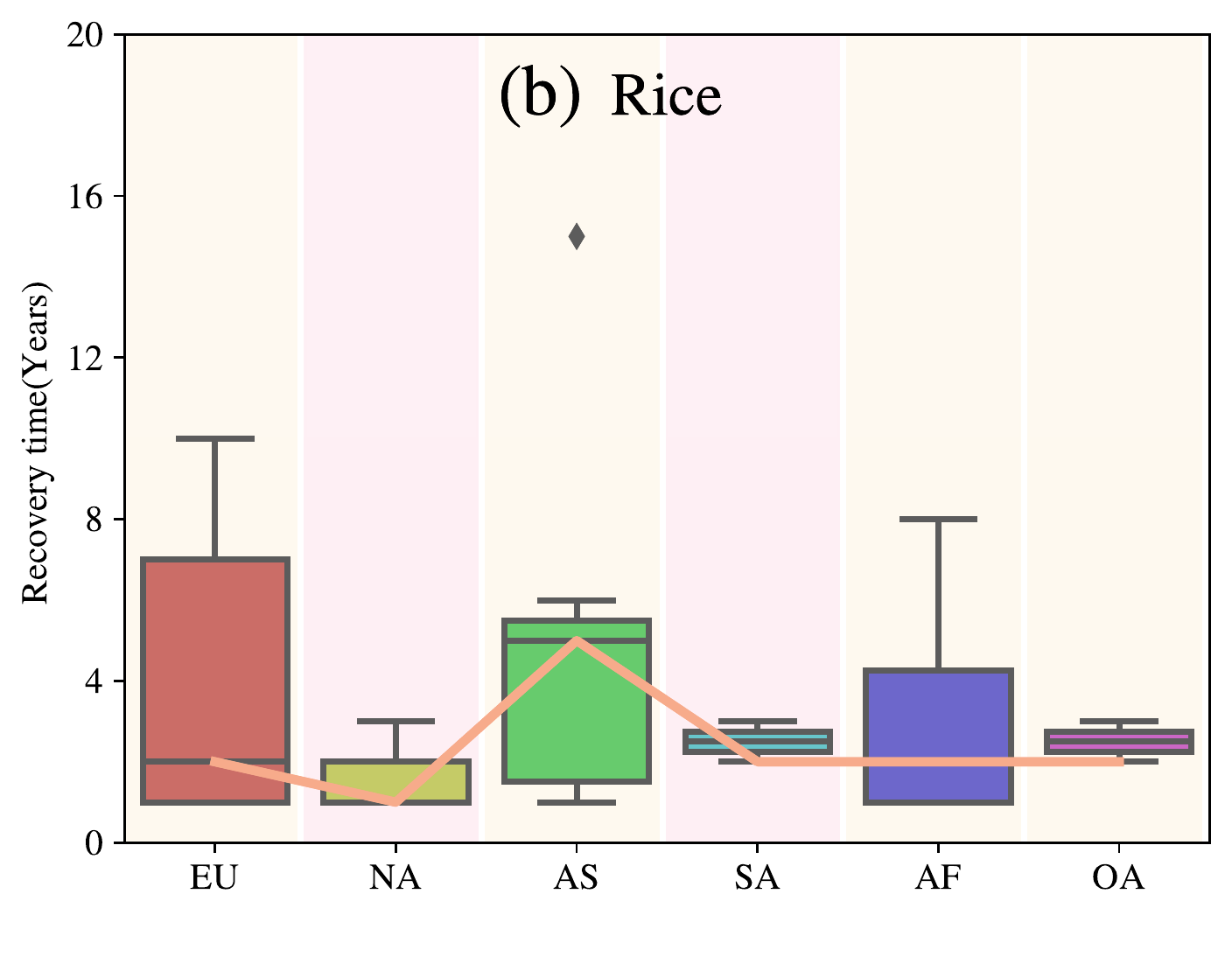}
    \includegraphics[width=0.475\linewidth]{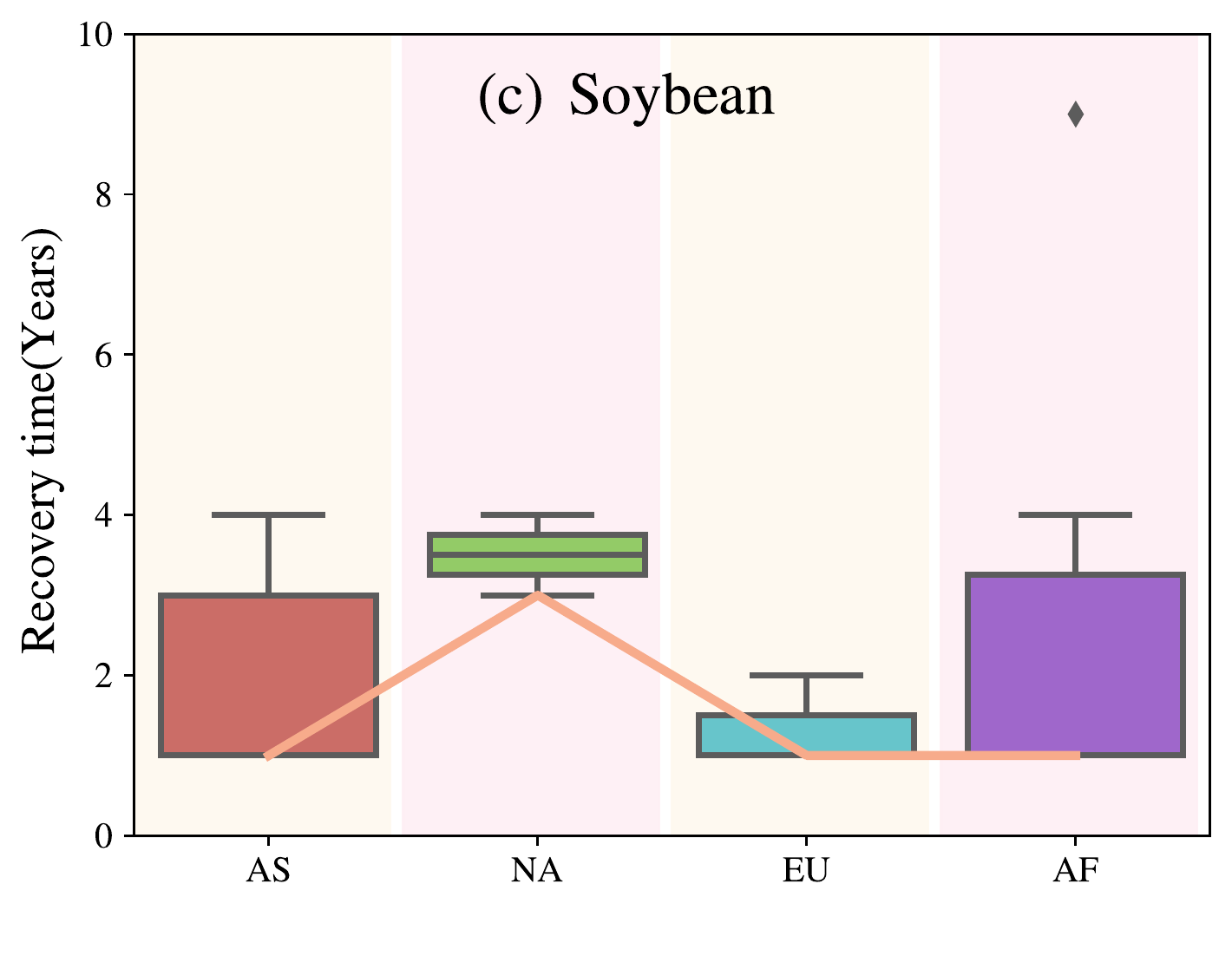}
    \includegraphics[width=0.475\linewidth]{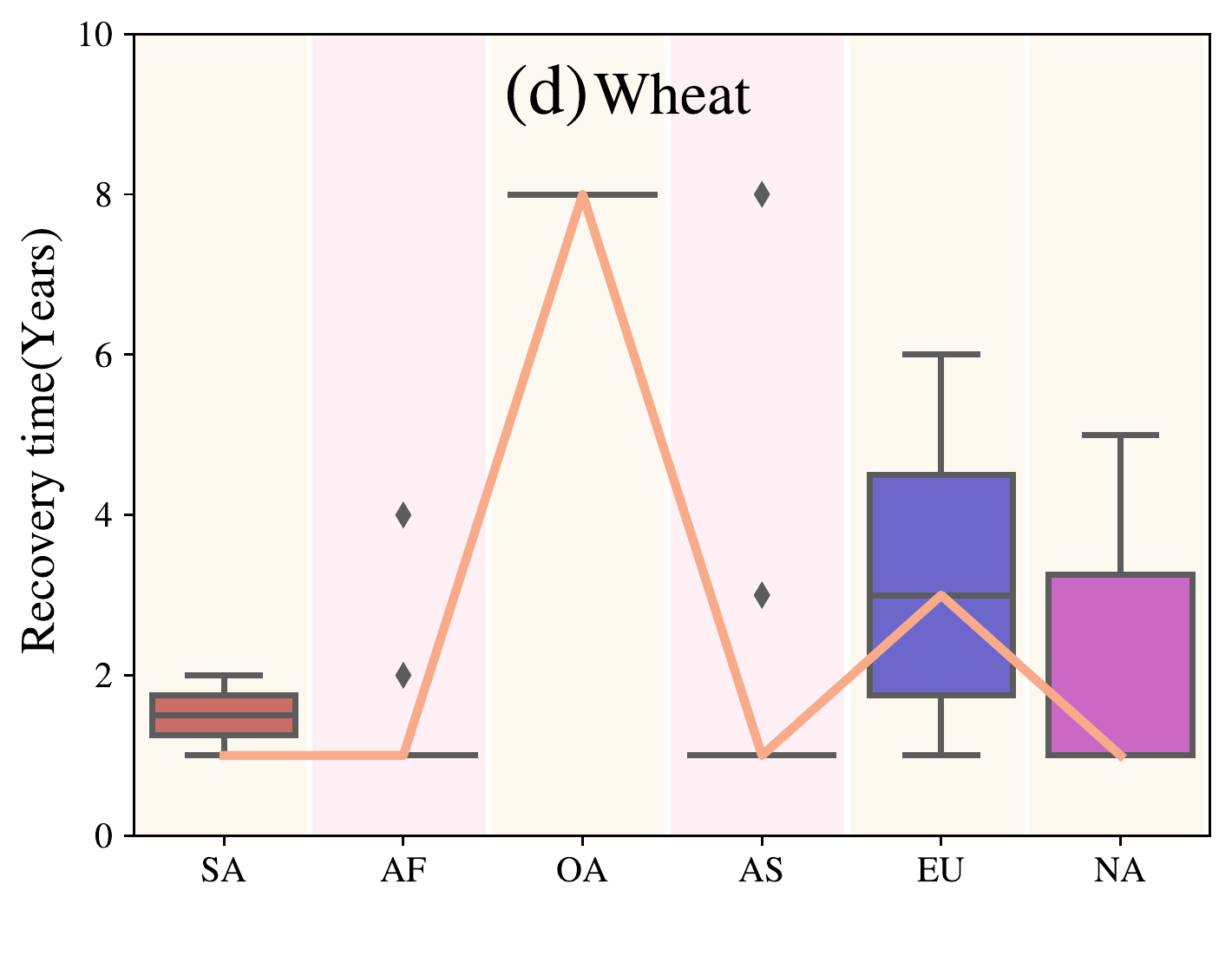}
    \caption{Recovery time of shocks by region for maize (a), rice (b), soybean (c) and wheat (d). Recovery was defined as returning to within 5\% of the previous 5-year average. The dark orange line shows the medians of the logarithmic shock magnitudes. Note that AF refers to Africa, AS refers to Asia, EU refers to Europe, OC refers to Oceania, NA refers to North America, SA refers to South America, and AN refers to Antarctica. We sort the regions according to shock rates.}
    \label{Fig:iCTN:shock:RecoveryTime}
\end{figure}

Figure~\ref{Fig:iCTN:shock:RecoveryTime} shows the recovery time of the crop import shocks in different regions from 1995 to 2018. Obviously, the recovery time varies across different regions and crops. In the case of maize, rice and wheat import shocks, most recoveries in Oceania were relatively slow with long recovery times. Most shocks in Asia recovered quickly, especially for maize, soybean and wheat. It may indicate that the crop import volumes in Asia will adjust in a short time. Asia is the major crop producer, but also relies on food imports to meet food supplies as growing population growth and biofuel demand. When food prices rise, most governments adjust food trade policies in a jiffy to stabilize domestic food markets \cite{Chang-Lee-Hsu-2013-PacRev}. Nevertheless, some shocks took a long time to recover. For example, the obvious outliers of shock recoveries in Africa shown in Fig.~\ref{Fig:iCTN:shock:RecoveryTime}(a) implies that the recovery time of some shocks was longer than 15 years and shocks were persistent. There are many outliers in Fig.~\ref{Fig:iCTN:shock:RecoveryTime}(d), indicating that wheat import shocks in some regions sometimes need a long time to recover. One possible implication of these results is related to the trade policies of some economies. For maize import shocks, recovery time fluctuated significantly. The mean recovery time was similar across regions except Oceania. The median recovery time of rice import shocks was shorter than six years. In Fig.~\ref{Fig:iCTN:shock:RecoveryTime}(c), the boxplot for South American is missing since shocks identified there did not recover 
throughout the study period. Shocks in other regions almost recovered in five years. In general, most recoveries in a region were quick when the shocks rates were not large.


\subsection{Results at the total food sector level}

\begin{figure}[h!]
        \subfigbottomskip=-1pt
    \subfigcapskip=-5pt
    \centering
    \subfigure[]{\label{level.sub.5}\includegraphics[width=0.39\linewidth]{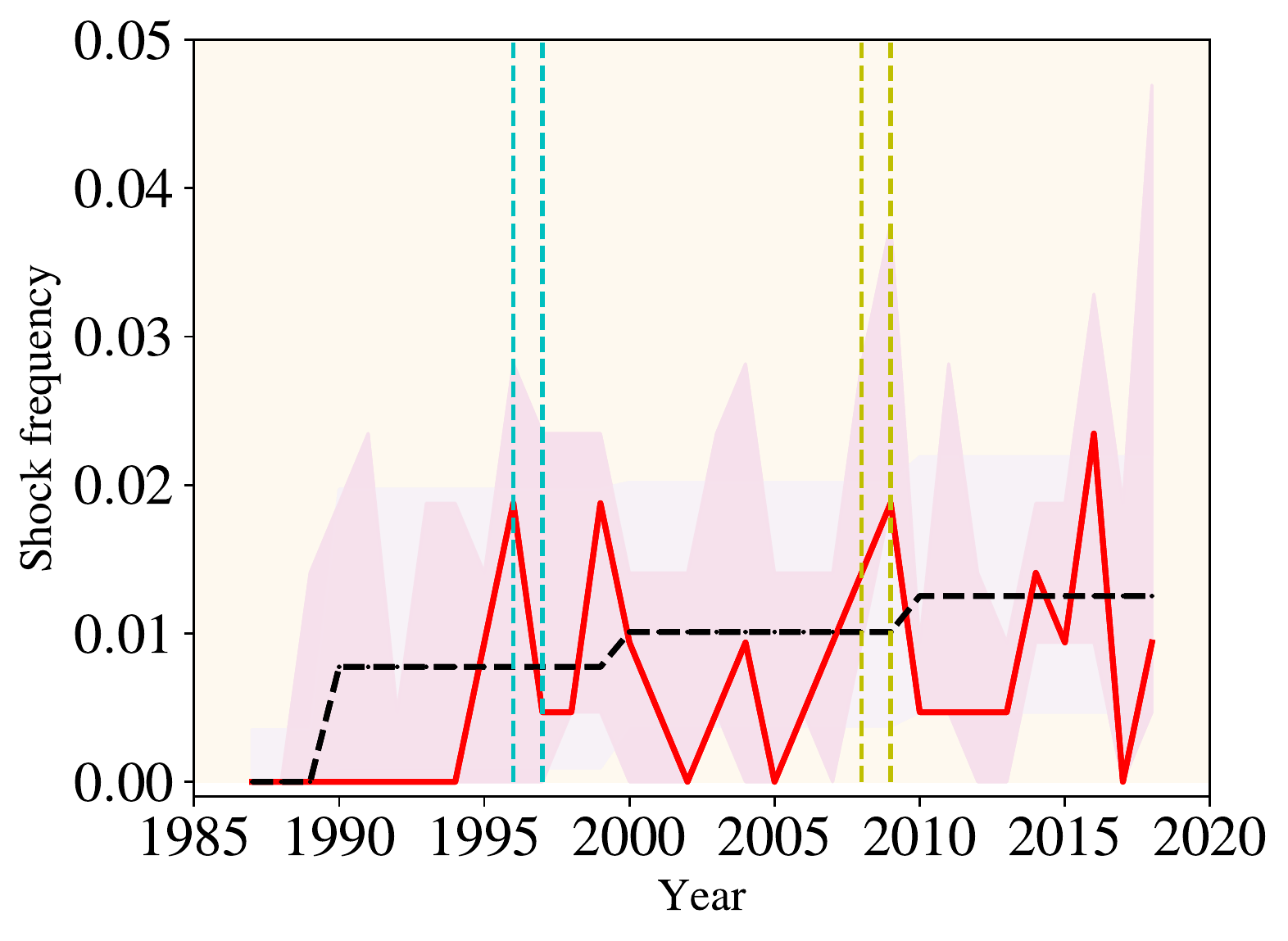}}
    \subfigure[]{\label{level.sub.6}\includegraphics[width=0.59\linewidth]{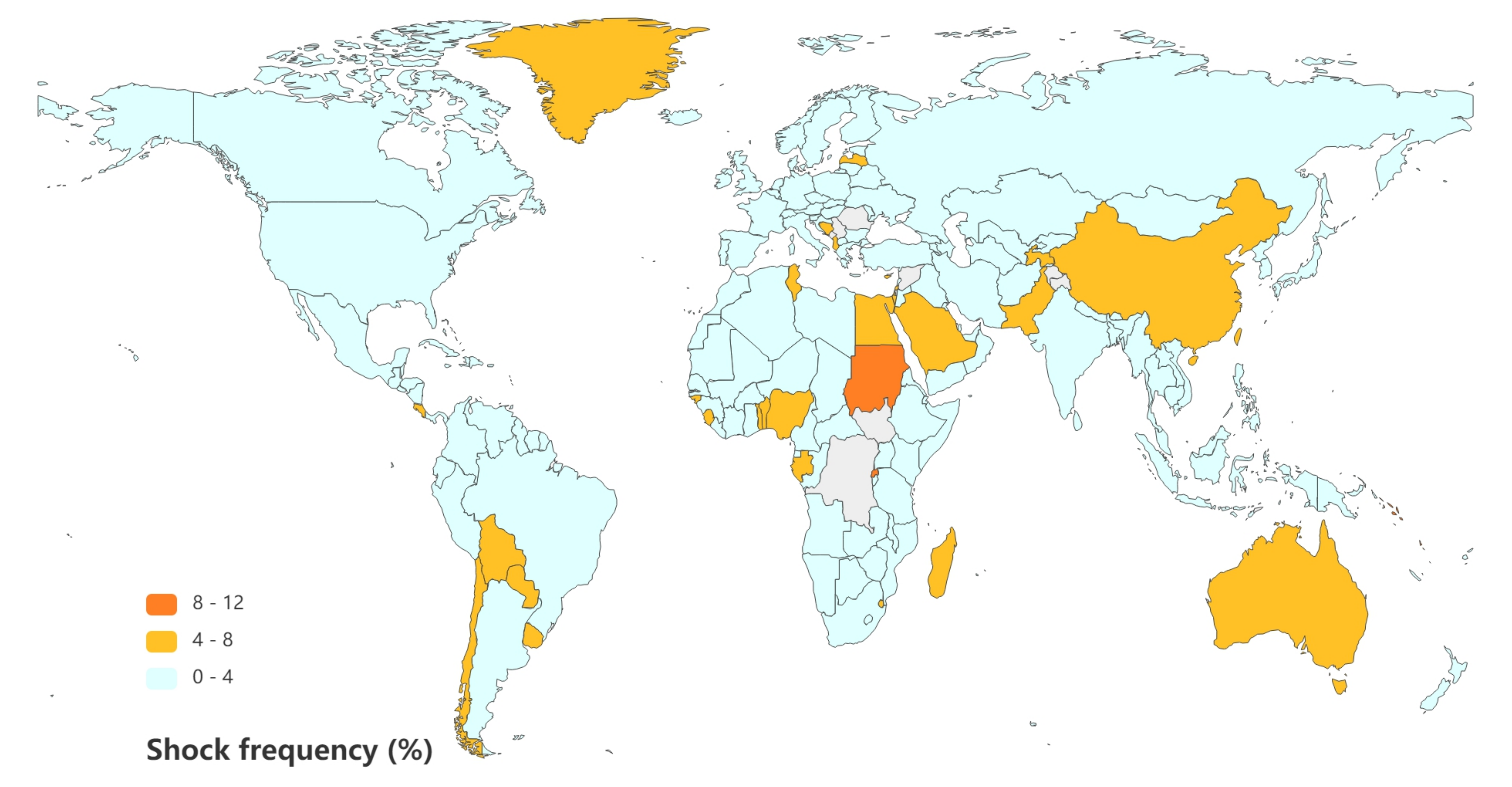}}\\
    \caption{Temporal trend (left panel) and spatial pattern (right panel) in crop supply shock frequency at the food sector level from 1995 to 2018. The events corresponding to the dotted blue and yellow lines are the financial crisis (1996/1997) and the food price crisis (2008/2009).}
    \label{Fig:iCTN:total:shock}
\end{figure}

The identification results of import shocks of individual crops are different from those of the aggregate of all crops. We detected 51 aggregate crop import shocks from 1995 to 2018. The trend of the decadal average shock frequency shown in Figure~\ref{Fig:iCTN:total:shock}(a)~(black dotted line) indicates that the shock frequency of aggregate crop import increased steadily. Especially, significant food import shocks (red line) occurred in 1996, 1999-2000, 2008-2009 and 2016. This coincided with the occurrence of the global food price crises, suggesting that the food import trade was affected by soaring food prices. Food price spikes would cause food import shocks in some economies. Additionally, we find that economies that have experienced a single crop import shock in a given year did not experience an aggregate food import shock. For example, Aruba had a maize import shock in 1999, but did not experience an aggregate food import shock. It may indicate that other food imports made up for the decrease in maize imports. Another example is that Bahamas had a rice import shock in 2010, but did not have an aggregate food import shock. The aggregate food import shock might be caused by a reduction in import trade volumes of a given crop, such as Eswatini. It had an aggregate food import shock because of a simultaneous maize import shock. The regional distribution of aggregate food import shocks is also different from that of single crop import shocks. It is obvious from Fig.~\ref{Fig:iCTN:total:shock}(b)~that the aggregate food import shocks mainly occurred in Africa, South America, Asia and Oceania. The shock rate of Sudan is the highest among all the economies.

\begin{figure}[h!]
    \centering
    \includegraphics[width=0.75\linewidth]{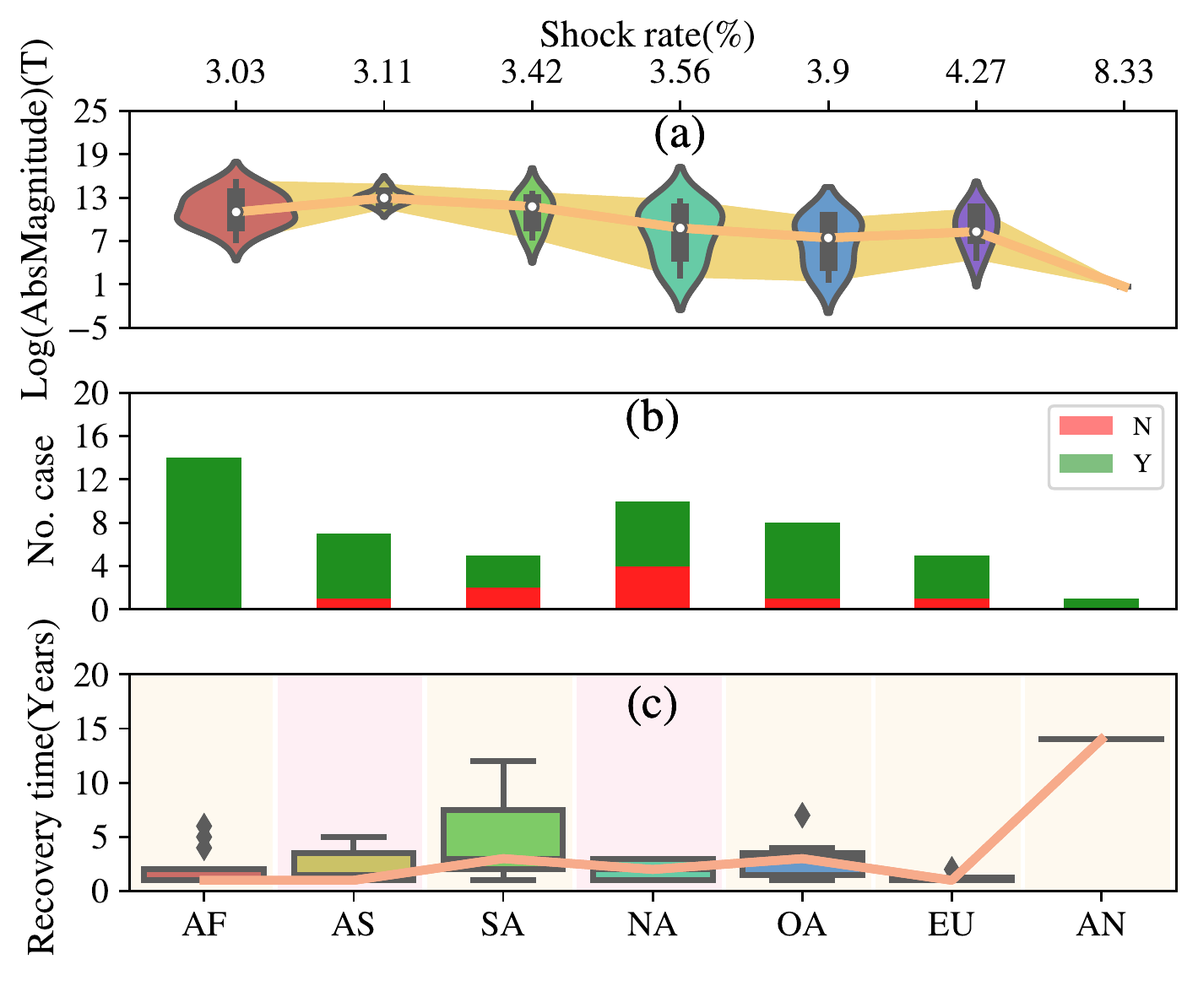}
    \caption{Shock rate (number of shocks divided by the number of time series in the region), magnitude, number of recovered and not recovered cases, and recovery time by region. Shocks were identified in international import volumes using the shock detection approach described in the methods. Recovery was defined as returning to within 5\% of the previous 5-year average. Note that AF refers to Africa, AS refers to the Asia, EU refers to Europe, OC refers to Oceania, NA refers to North America, SA refers to South America, and AN refers to Antarctica. N refers to that the shock was not recovered while Y refers to that the shock was recovered. We sort the regions according to shock rates.}
    \label{Fig:iCTN:total:shock:region}
\end{figure}

Figure~\ref{Fig:iCTN:total:shock:region} shows the magnitude of aggregate food import shocks, the number of recovered and not recovered cases, and the recovery time in the six continents from 1995 to 2018. It can be seen from Fig.~\ref{Fig:iCTN:total:shock:region}(a) that the shock magnitude varies in different regions. The shock magnitude in Asia was the largest, followed by Africa. Fig.~\ref{Fig:iCTN:total:shock:region}(b) is a histogram of the number of recovered and not recovered cases. Shocks in Africa have all returned to the corresponding pre-shock levels, while shocks elsewhere recovered partly, particularly in North America and South America. As can be seen from Fig.~\ref{Fig:iCTN:total:shock:region}(c), except in South America where the shock recovery time was relatively long, the average shock recovery time in Asia, North America and Europe was less than 5 years and that in Africa and Oceania was more than 5 years.

These results reveal that the structure and characteristics of the single and aggregate crop import trade are significantly different. When studying the international food trade networks, we cannot simply aggregate different kinds of crops to analyze the total food import trade. Nevertheless, there are also similarities between aggregate and single food import shocks. One is that the shock frequencies were both higher in 2008-2009. It was mainly affected by the global food price crisis in 2007/2008. Another point is that between 1995 and 2018, Africa and Asia both experienced large shocks but recovered quickly.

\subsection{Driving factors}

Our results indicate that some internal or external factors prompt economies to have food import shocks. Since it is not the main topic, we do not analyze the drivers of shocks in depth. We only try to preliminarily and qualitatively analyze how the import diversity, the food import dependency ratio and the food price affect the shock rate and the shock recovery. As can be seen from Table~\ref{tbl:Correlation}, there is seemingly a weak negative correlation between the import diversity and the shock rate of economies. When the import diversity of an economy is high, import trade shocks are less likely to occur. There is a weak positive correlation between the food import dependency ratio and the import trade shock rate. When the food supply of an economy is more dependent on food imports, import shocks are more likely to occur. On the contrary, it can be seen from the Table~\ref{tbl:Correlation} that the proportion of recovered shock has a slight positive correlation with the import diversity of an economy, and a negative correlation with the food import dependency ratio. High import diversity and low import dependency ratio lead to high recovery rate of import shocks.

\begin{table*}[!h]
\begin{center}
\setlength{\tabcolsep}{12pt}
\caption{Relationships between the shock rate and the import diversity, the shocks rate and the food import dependency ratio, the recovery rate and the import diversity, and the recovery rate and the food import dependency ratio. The recovery rate means the ratio of the number of recovered cases to the number of shocks from 1995 to 2018.}
\medskip
\begin{tabular}{@{}lll@{}} 
\toprule
 Pearson correlation coefficient  &   Diversity & Import dependency ratio \\
 \midrule
     Shock~rate & -0.00632   & ~~~~~~~~0.03807
 \\
     Recovery~rate & 0.11430  & ~~~~~~~-0.33482
\\
    \bottomrule
  \end{tabular}
  \label{tbl:Correlation}
\end{center}
\end{table*}

\begin{figure}[h!]
\centering
    {\includegraphics[width=0.53\linewidth]{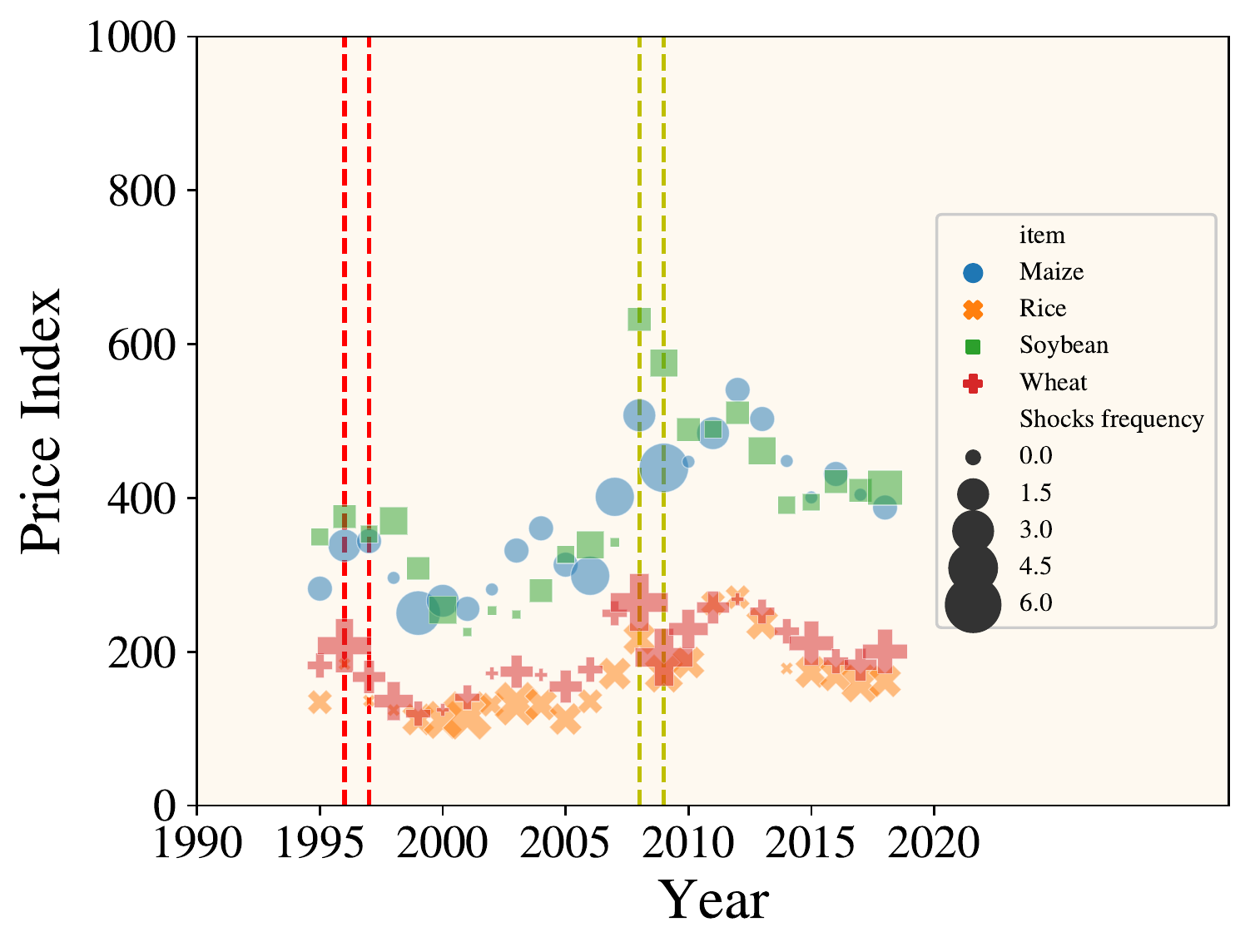}}
    \caption{The relationship between food import shock rate and food price. The abscissa is the time and the ordinate is the crop price index. The colors of different kinds of crops are different, and the size of the point represents the value of the shock rate. The events corresponding to the dotted red and yellow lines are the financial crisis (1996/1997) and the food price crisis (2008/2009).}
    \label{Fig:iCTN:shock:price}
\end{figure}

In comparison, food prices seem to have a more striking impact on the shock rate. 
We characterized the temporal trend of the four food price indicators and the shock rates from 1995 to 2018 in  Fig.~\ref{Fig:iCTN:shock:price}, and found that in 2008-2009, both food prices and food import shock frequencies were high. However, we only simply analyzed the relationship between the temporal trend of these two variables, rather than analyzing quantitatively the intrinsic causality. Despite its exploratory 
essence, this study offers some insights into the drivers of food import shocks.

\section{Discussion and conclusion}
\label{S4:Discussion}

Ensuring a stable food supply is key to food security. Some economies cannot meet their population's food needs through domestic production due to resource, land and climate limitations. Therefore, food import trade is an important approach to ensure food supplies in food-deficient economies. However, food import trade is easily affected by the food production and trade policies of major exporters. It may cause a food import trade shock and a food supply risk in importing economies. Shocks to food import trade pose significant hazards to improving global food security, and other sustainability targets. Therefore, we constructed the international food trade network and identified outliers of import volumes based on locally weighted polynomial regression. This helps to detect food import trade shocks, understand the external food supply risk of economies, and measure food security from the dimension of food availability.

We investigated historical global trends of food import shocks across maize, rice, soybean and wheat from 1995–2018. We used an established, standardized approach to identify shocks in food import volumes taken from FAO. Adapting locally weighted polynomial regression models, we detected shocks through breaks in the autocorrelation structure of a time series and integrated shock points with some special events. Here, we mapped food import shock frequency, compared the temporal evolution trend and spatial distribution of food import shocks across different crops and analyzed the shock intensity and shock recovery across different regions. One interesting finding is that as the international food trade growing (see Fig.~\ref{Fig:iCTN:sumW:t:1986-2018}), shocks did not become more frequent (see Fig.~\ref{Fig:iCTN:ShockFreq:4Crops}) over time. It suggests that shocks are a common feature of these international food trade systems. The temporal trend of the import shocks of four crops had their specific characteristics, but had similarities in some years. In particular, during 2008-2009, all kinds of crop import trade had obvious shocks. This also accords with the food crisis caused by soaring global food prices. It suggests an evident link between food imports and food prices. Price spikes would lead to significant reductions in the external food supply in some economies.

The import shock rates vary across different crops and regions. We find that North America, Africa and Asia have experienced more frequent crop import shocks. Spatial features of import shocks depend on the types of crop. Rice and wheat import shocks occurred more frequently and more widely. These shocks mainly occurred in Africa, Asia and North America. Maize import shocks mainly occurred in Africa and North America, while soybean import shocks were mainly in Asia and Europe. Since regions proposed different food trade strategies according to their local crop production and food demand, crop import shock rates exhibited spatial features. Rice is the staple food feeding more than half of the world's population and has a major impact on global food security. Nearly 40\% of rice consumption in Africa comes from imports, especially sub-Saharan Africa, which is a net importer of rice \cite{Seck-Tollens-Wopereis-Diagne-Bamba-2010-FoodPolicy}. Before 2000, the global rice supply exceeded demand and market prices were stable because of favorable weather and overproduction. At that time, the import supply of rice in some economies was fully guaranteed. After then, the external rice demand increased as the boost in rice yield declined. The situation worsened by the end of 2007, when many rice exporters reduced exports or stopped supplying rice, leading to a global rice crisis \cite{Seck-Diagne-Mohanty-Wopereis-2012-FoodSecur}. This is a big shock to major rice importers such as those in Africa. Therefore, African economies had frequent rice import shocks after 2000. In addition to the shock frequency, the shock intensity (or shock magnitude) is also an important indicator to describe the food import shock. In this paper, the shock intensity is measured by calculating the difference between the under-shock import volumes and the five-year average pre-shock import volumes. It is found that the shock magnitude in South America was high. However, the shock magnitudes in North America, Europe and Africa were related to the crop variety. At the same time, we noted that the shock intensity is not positively correlated with the shock frequency. The import shock rate in South America was low, but the shock intensity was large, opposite to the case in North America. This suggests that regions with frequent shocks are not necessarily exposed to severe import trade shocks, which need to take into account the local agricultural food production and trade policies.

The food system has the capacity to provide sufficient and accessible food, in the face of various and even unforeseen shocks. Some shocks are temporary and recover in a short time. However, some shocks are persistent. We take the five-year average food supply before the shock as the reference. When the import volumes return to 95\% of the pre-shock levels, we believe that the system has recovered from the shock. As can be seen from the results, there are significant indigenous differences in shock recoveries across regions. Although the shock rate in Asia was high, most shocks have recovered before long. It indicates that Asia can adjust imports in a timely manner and return to normal external food supply after a substantial reduction in food imports. The food security situation in Asia, particularly in East Asia, is better than the rest of the world. Food security in East Asia is mainly guaranteed by domestic production. Although after 2000, in order to meet the growing population demand, Asian food imports have doubled, but Asia is still less dependent on food imports. However, rising energy costs and food prices due to increased food demand for biofuels have exacerbated food shortages in parts of Asia. Most government interventions focus on short-term policies. They prefer to stabilize domestic food prices by adopting temporary trade restrictions or price controls \cite{Chang-Lee-Hsu-2013-PacRev}. These measures may lead to a reduction in food import trade within a short time, causing an import trade shock. But the situation can be adjusted promptly as the food market situation changes.

Shocks in some regions did not recover quickly. Most of the soybean import shocks in Europe and North America have yet to recover. Before 2015, Europe and the EU in particular faced severe soybean deficits. Soybean imports account for 95\% of European annual soybean consumption. The accession of economies along the Danube river and the association treaties signed by the European Union and other Eastern European economies have created new opportunities for increasing soybean production. These regions plant new soybean varieties with high yields and excellent traits genetics by developing or improving the cropping technologies for soybeans \cite{Dima-2015-AgricAgricSciProc}. When an economy's domestic soybean production increases, it would become less dependent on external imports. It causes long-lasting import shocks. This is seemingly able to explain why most soybean import shocks could not recover.

If the food import shock can recover in a short time, the external food supply of the economy will not be significantly affected. But if the shock recovery time is more than 15 years, the external food supply declines in a long time. The economy is prone to suffer food scarcity unless it improves domestic food production to meet internal supplies.

We further analyzed the impacts of import diversity, import dependency ratio, and crop price on shock rate and shock recovery rate. It is unveiled that high import diversity and low import dependency ratio leads to low shock rate and high recovery rate. In addition, high crop prices are often followed by high shock rates.

This paper sets out to detect crop import shocks based on the method of outlier identification. Since shocks represent sudden drops in import volumes, the detection approach does not identify long-term, more gradual reductions. Further, this method does not identify shocks in systems with high variability. In systems with high variability, large deviations are frequent and are therefore not shocks considered in this analysis. Our study is limited by the lack of information on factors causing import shocks, but has suggested that the food import trade risks involved. In addition, this study contributes to our understanding of the temporal and spatial features of crop import trade shocks. It will be of interest to focus on the impact of the food crisis and the trade structure on the international food import trade. The insights gained from this study may be of assistance in improving our understanding of the external food supply risks and paying more attention to the availability of food. Considerably more work needs to be done to conduct a more comprehensive and systematic analysis of food supply issues in the context of food trade, production, consumption and reserves \cite{Marchand-Carr-Dell'Angelo-Fader-Gephart-Kummu-Magliocca-Porkka-Puma-Ratajczak-Rulli-Seekell-Suweis-Tavoni-D'Odorico-2016-EnvironResLett}. This would be a fruitful area for further work, which assists in our broader understanding of food security.

\section*{Conflict of Interest Statement}








\section*{Data Availability Statement}
Publicly available datasets were analyzed in this study. This data can be found here:{{ https://www.fao.org}}.


\end{document}